\documentclass[conference]{IEEEtran}
\IEEEoverridecommandlockouts
\usepackage{cite}
\usepackage{amsmath,amssymb,amsfonts}
\usepackage{algorithmic}
\usepackage{algorithm}
\usepackage{graphicx}
\usepackage{textcomp}
\usepackage{xcolor}
\usepackage{url}
\usepackage{enumitem}
\usepackage{multirow}
\usepackage{booktabs}
\usepackage{graphicx}
\usepackage{algorithmic}
\usepackage{placeins}
\usepackage{mathtools}
\usepackage{bm}
\usepackage{diagbox}
\usepackage{makecell}
\usepackage{subfigure}
\usepackage{subcaption}

\usepackage{lipsum,subeqnarray}

\DeclareMathOperator{\diag}{diag}

\newcommand{\cc}{\mathbf{c}}
\newcommand{\nn}{\mathbf{n}}

\newcommand{\x}{\mathbf{x}}

\newcommand{\xx}{\mathbf{\tilde{x}}}
\newcommand{\xxx}{\mathbf{\hat{x}}}

\newcommand{\yyy}{\mathbf{\hat{y}}}
\newcommand{\uu}{\bm{u}}
\newcommand{\zz}{\bm{z}}
\newcommand{\vv}{\bm{v}}

\newcommand{\p}{\mathbf{P}}

\newcommand{\Ex}{\mathbb{E}}
\newcommand{\pr}{\mathbb{P}}

\newcommand{\rt}{\right}
\newcommand{\lt}{\left}

\newcommand{\I}{\mathbf{I}}
\newcommand{\A}{\mathbf{A}}
\newcommand{\B}{\mathbf{B}}
\newcommand{\M}{\mathbf{M}}

\newcommand{\W}{\mathbf{W}}
\newcommand{\D}{\mathbf{D}}

\newcommand{\LL}{\mathbf{L}}

\newcommand{\Rn}{\mathbb{R}}

\DeclareMathOperator{\Vol}{vol}

\begin{document}

\title{Effective Delayed Patching for Transient Malware Control on Networks}

\author{\IEEEauthorblockN{Minh Phu Vuong}
\IEEEauthorblockA{\textit{Texas State University}\\
cty13@txstate.edu}
\and
\IEEEauthorblockN{Chul-Ho Lee}
\IEEEauthorblockA{\textit{Texas State University}\\
chulho.lee@txstate.edu}
\and
\IEEEauthorblockN{Do Young Eun}
\IEEEauthorblockA{\textit{North Carolina State University}\\
dyeun@ncsu.edu}
}

\maketitle

\begin{abstract}
Patching nodes is an effective network defense strategy for malware control at early stages, and its performance is primarily dependent on how accurately the infection propagation is characterized. In this paper, we aim to design a novel patching policy based on the susceptible-infected epidemic network model by incorporating the influence of patching delay--the type of delay that has been largely overlooked in designing patching policies in the literature, while being prevalent in practice. We first identify ‘critical edges’ that form a boundary to separate the most likely infected nodes from the nodes which would still remain healthy after the patching delay. We next leverage the critical edges to determine which nodes to be patched in light of limited patching resources at early stages. To this end, we formulate a constrained graph partitioning problem and use its solution to identify a set of nodes to patch or vaccinate under the limited resources, to effectively prevent malware propagation from getting through the healthy region. We numerically validate that our patching policy significantly outperforms other baseline policies in protecting the healthy nodes under limited patching resources and in the presence of patching delay. 
\end{abstract}

\begin{IEEEkeywords}
Epidemic modeling and control, malware propagation, patching delay, graph partitioning
\end{IEEEkeywords}

\section{Introduction}
Recent technological advancements have led to a rapid increase in the number of devices connected to the Internet, including conventional personal computers and various mobile and IoT devices, which form large complex networks. End-users, through these connected devices, share a massive amount of information across different networks, making them vulnerable to cyber attacks and malware infections. Once successfully installed, the malware propagates in a network by exploiting devices' vulnerabilities, and it can infect a large number of devices in a short time~\cite{durumeric2014matter, trendmicro2022zerologon, microsoft2022springshell, kaseya2022incident, zetter2014stuxnet}. A typical example is the Code Red 2 variant that infected 350,000 machines within a span of 24 hours~\cite{zou2002code}.

To control the spread of malware, we need to characterize its spreading dynamics and devise the corresponding combating strategies. Common models to characterize the spread of malware infections are the susceptible-infected-susceptible (SIS) model, the susceptible-infected-recovered (SIR) model, and their extensions~\cite{pei2025unraveling, nowzari2016analysis, van2008virus, newman2010networks}. Most studies based on these models largely focus on their steady-state behaviors (where the cure/vaccine is readily available for all nodes) to establish the ``epidemic threshold" under which the epidemic dies out eventually over time. The countermeasures developed based on these models thus attempt to restrict the infection below the threshold by increasing the recovery rate or modifying the network connectivity to curtail the spread. 

However, a cure for malware (also called `patching') is not readily available upon a malware attack, as it may take days or weeks to identify an exploit, let alone its cure. In this transient period, we can mitigate the spread by deploying (temporary) patches relying on advanced security measures such as threat prevention and deep learning capabilities \cite{klarich2022inline, cynet2022zeroday, checkpoint2022zeroday}; however, these patches are expensive and certainly not available at scale. Therefore, at the early stage of the spread, the aforementioned models that consider a specific recovery rate for each node and the corresponding combating strategies are no longer valid. A more natural fit here is the susceptible-infected (SI) model, which assumes that a node, once infected, will stay infected, thus better representing the transient behavior of malware propagation. This observation led to our earlier work~\cite{lee2019transient} on characterizing the transient dynamics of the SI model on a network and establishing a tighter upper bound on the likelihood of each node being infected at any time $t$.

\vspace{2pt}
\textbf{Motivation:} In addition to the delay caused by patching development, there is another equally important delay that exists when we actually implement patches to nodes or devices. This delay has been clearly seen in various security patches practically implemented over the years \cite{durumeric2014matter, arghire2019spectre, umich2016drown}. For instance, the Heart Bleed vulnerability allows the attackers to read sensitive memory (containing cryptographic keys) of vulnerable servers \cite{durumeric2014matter}, and its mitigation requires an administrator to replace the cryptographic keys, revoke the compromised certificate, and apply the security patch. These actions account for a significant amount of time. Other similar examples are Spectre and Meltdown security vulnerabilities and DROWN attack \cite{arghire2019spectre,umich2016drown}, which also require many time-consuming actions to execute patching. A study conducted by tCell~\cite{rapid72018appsec} shows that the mean time to patch critical vulnerabilities takes close to 38 days. Google's Project Zero~\cite{hawkes2018projectzero} also reports that it takes 15 days on average for vendors to patch a vulnerability that is being used in active attacks.

Ideally, the strategies for selecting which nodes to patch should incorporate the patching delay $T$ by analyzing the time-dependent influence of malware propagation on the network; however, this delay has been largely ignored in the literature. For instance, it is assumed in \cite{lee2019transient} that the patching for nodes marked for vaccination becomes effective immediately (i.e., $T \!=\! 0$), creating issues in real-world scenarios where it takes a non-negligible amount of time to take into effect. While our recent work~\cite{Li23Networking} considers the presence of this patching delay, it is only limited to \emph{tree} topologies. Therefore, it is crucial to address this gap by designing a new patching policy that takes into account the delay $T$ and efficiently determines which nodes to patch on \emph{general} graphs. 

\vspace{2pt}
\textbf{Our Contributions:} In this work, we present a novel mathematical framework based on the SI model, which leads to a new patching policy that efficiently selects the most feasible nodes for patching under a limited budget. With the influence of time delay $T$ incorporated into the patching policy, our node selection traces the state of infection more accurately, where eligible candidates continuously evolve within a network as the infection is actively spreading in the meantime. The main contributions of this work are as follows:

\begin{itemize}[itemsep=2pt,leftmargin=1.3em,topsep=2pt]
    \item We highlight the presence of patching delay $T$ in real-world scenarios and illustrate its importance in effectively mitigating malware propagation. 
    \item To predict the spread of malware infections and leverage such information, we encode the infection probability of each node at time $T$ into the graph model as `edge weights', leading to the notion of `critical edges' with maximal weights across the boundary between healthy and infected groups of nodes.
    \item We formulate a constrained graph partitioning problem to identify the boundary that separates the most likely infected nodes from the nodes which would still be healthy after $T$. We then leverage its solution to determine which nodes to patch or vaccinate under the budget constraint.
    \item We numerically validate that our policy significantly outperforms other baseline policies, including the reactive policy introduced in \cite{lee2019transient}, in protecting healthy nodes under the same budget constraint.
\end{itemize}

The rest of the paper is organized as follows. In Section~\ref{sec:prelim}, we review the SI epidemic model and its approximations and demonstrate the importance of patching delays. Section~\ref{sec:framework} presents the details of our proposed mathematical framework that formulates a constrained graph partitioning problem to mitigate malware propagation while accounting for patching delay $T$ under a limited budget. We present numerical simulation results in Section~\ref{sec:simulation} and conclude in Section~\ref{sec:conclusion}.

\section{Preliminaries}\label{sec:prelim}
\subsection{Notations}

We present notations that will be used throughout the paper. For any two column vectors $\uu \!=\! [u_1, u_2, \ldots, u_n]^T \!\in\! \Rn^n$ and $\vv \!=\! [v_1, v_2, \ldots, v_n]^T \!\in\! \Rn^n$, we write $\uu \preceq \vv$ if $u_i \leq v_i$ for all $i = 1,2 \ldots, n$. Let $\bm{1}$ and $\bm{0}$ denote the $n$-dimensional all-one and all-zero column vectors, respectively. For a function $f$: $\Rn \!\to\! \Rn$ and for a column vector $\uu \!\in\! \Rn^n$, we write $f(\uu)$ as an $n$-dimensional column vector with elements $f(u_i)$. Similarly, we write $\diag(\uu)$ to be an $n \times n$ diagonal matrix with diagonal entries $u_i$. Let $\I \!=\! \diag(\bm{1})$ be an $n \times n$ identity matrix. 

\subsection{SI Epidemic Model and Its Approximations}

Consider a connected, undirected graph $G \!=\! (N, E)$ as a network model, where $N \!=\! \{1, 2, \ldots , n\}$ is a set of nodes and $E$ is a set of edges, indicating neighboring relationships between nodes. The graph $G$ is characterized by an $n \times n$ adjacency matrix $\A \!=\! [a_{ij}]$ with elements $a_{ij} \!=\! 1$ if there is an edge between nodes $i$ and $j$, i.e., $(i,j) \!\in\! E$, and $a_{ij} \!=\! 0$ if otherwise. In the SI model on a graph $G$, the infection process is primarily governed by the network connectivity, i.e., if node $i$ is infected and node $j$ is susceptible for $(i,j) \!\in\! E$, then $i$ can infect $j$ with a common infection rate $\beta > 0$. 

Let $X_i(t) \!\in\! \{0,1\}$ denote the state of node $i$ at time $t$, where $X_i(t) \!=\! 1$ indicates that node $i$ is infected at $t$, and $X_i(t) \!=\! 0$ indicates that node $i$ is healthy and susceptible to infection at $t$. We define $x_i(t) \!\triangleq\! \pr\{X_i(t) \!=\! 1\} \!=\! \Ex\{X_i(t)\} \!\in\! [0,1]$ to be the probability that node $i$ is infected at time $t$. In other words, node $i$ is healthy at time $t$ with probability $1 \!-\! x_i(t)$. Letting $\x(t) \!\triangleq\! [x_1(t), x_2(t), \ldots, x_n(t)]^T$ be a column vector with elements $x_i(t)$ at time $t$, we can write the following ODE for the SI model: For any node $i \in N$,
\begin{equation}
  \frac{d x_i(t)}{dt} = \beta (1-x_i(t)) \sum_{j\in N} a_{ij} x_j(t), \quad t \geq 0, \label{basic-SI}
\end{equation}
with an initial condition $\x(0)$.

Despite its simple form, the coupled, nonlinear ODE in (\ref{basic-SI}) cannot be solved in a closed form for most cases. Thus, a linear approximation has been adopted in the literature to obtain an approximate solution to (\ref{basic-SI})~\cite{Canright06,newman2010networks,mei2017dynamics}. Specifically, since $1 - x_i(t) \leq 1$, we have, for any $i \in N$ and for any time $t \geq 0$,
\begin{equation*}
  \frac{d x_i(t)}{dt} =  \beta (1 - x_i(t)) \sum_{j\in N} a_{ij} x_j(t)   \leq \beta \sum_{j\in N} a_{ij} x_j(t). \label{SI-inequality}
\end{equation*}
Then, letting $\xx(t) \triangleq [\tilde{x}_1(t), \tilde{x}_2(t), \ldots, \tilde{x}_n(t)]^T$ be the solution to the linear dynamical system in the upper bound, we have 
\begin{equation}
  \x(t) \preceq \xx(t) = e^{\beta  t \A}\x(0), \quad t \geq 0, \label{basic-SI-linear-solution}
\end{equation}
provided that $\xx(0) \!=\! \x(0)$. This upper bound works only for small values of $t$ and when $\x(0) \approx \bm{0}$, and becomes useless even for moderate values of $t > 0$ as it grows exponentially, contradicting that each $x_i(t) \in [0, 1]$ is a probability~\cite{lee2019transient}.

\begin{figure*}[t!]
    \centering
    \subfigure[$t=0$]{%
        \includegraphics[width=0.31\linewidth]{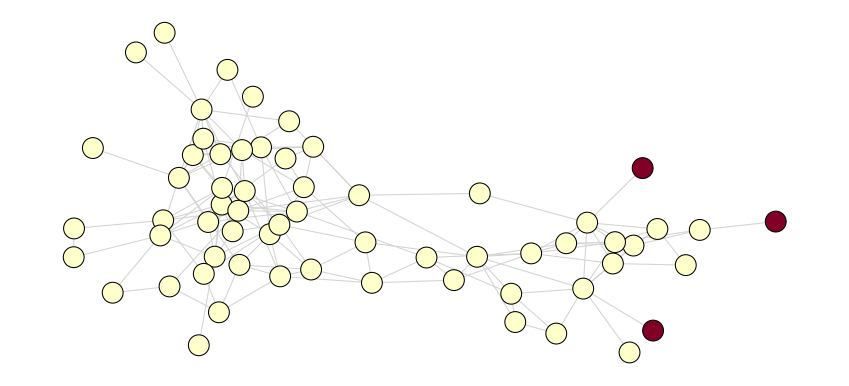}
    }
    \subfigure[$t=10$]{%
        \includegraphics[width=0.31\linewidth]{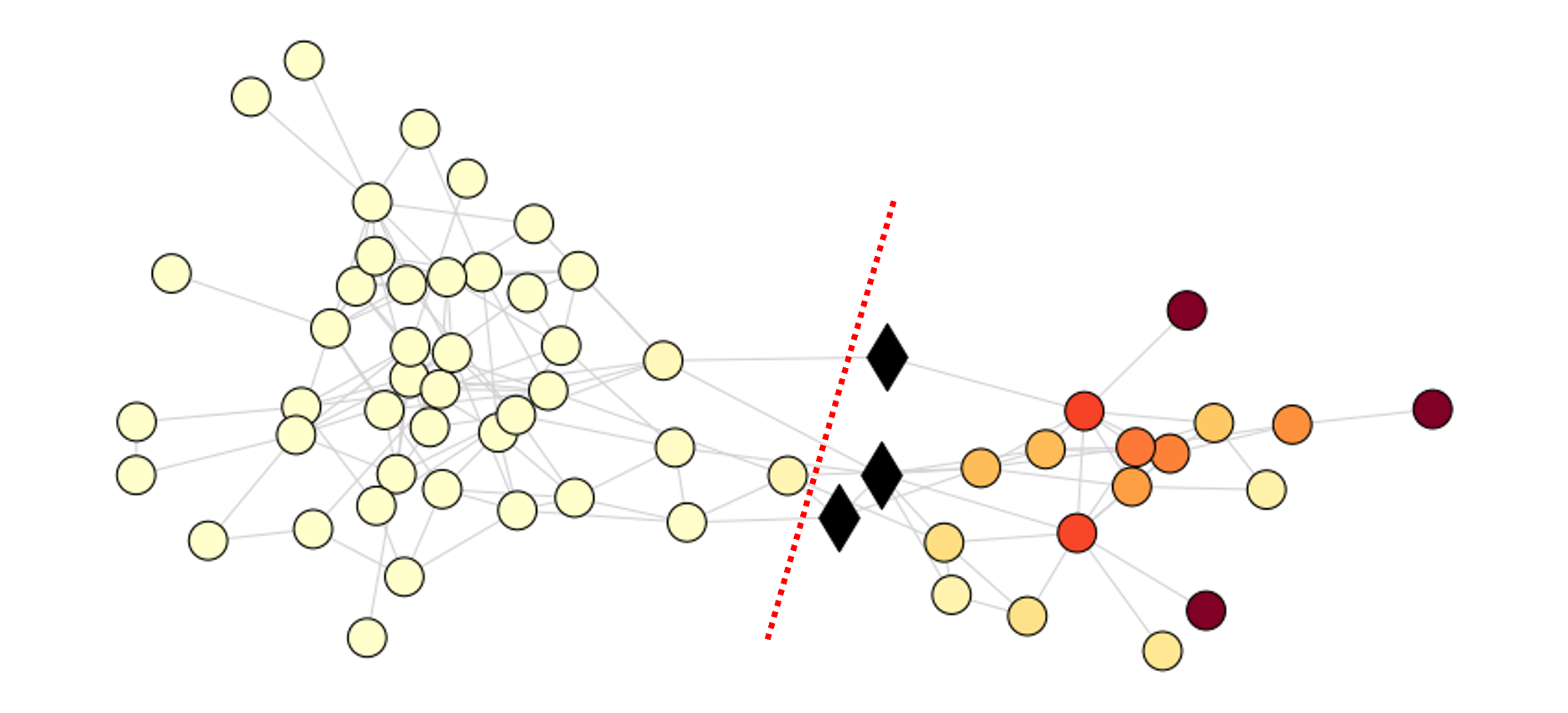}
    }
    \subfigure[$t=15$]{%
        \includegraphics[width=0.31\linewidth]{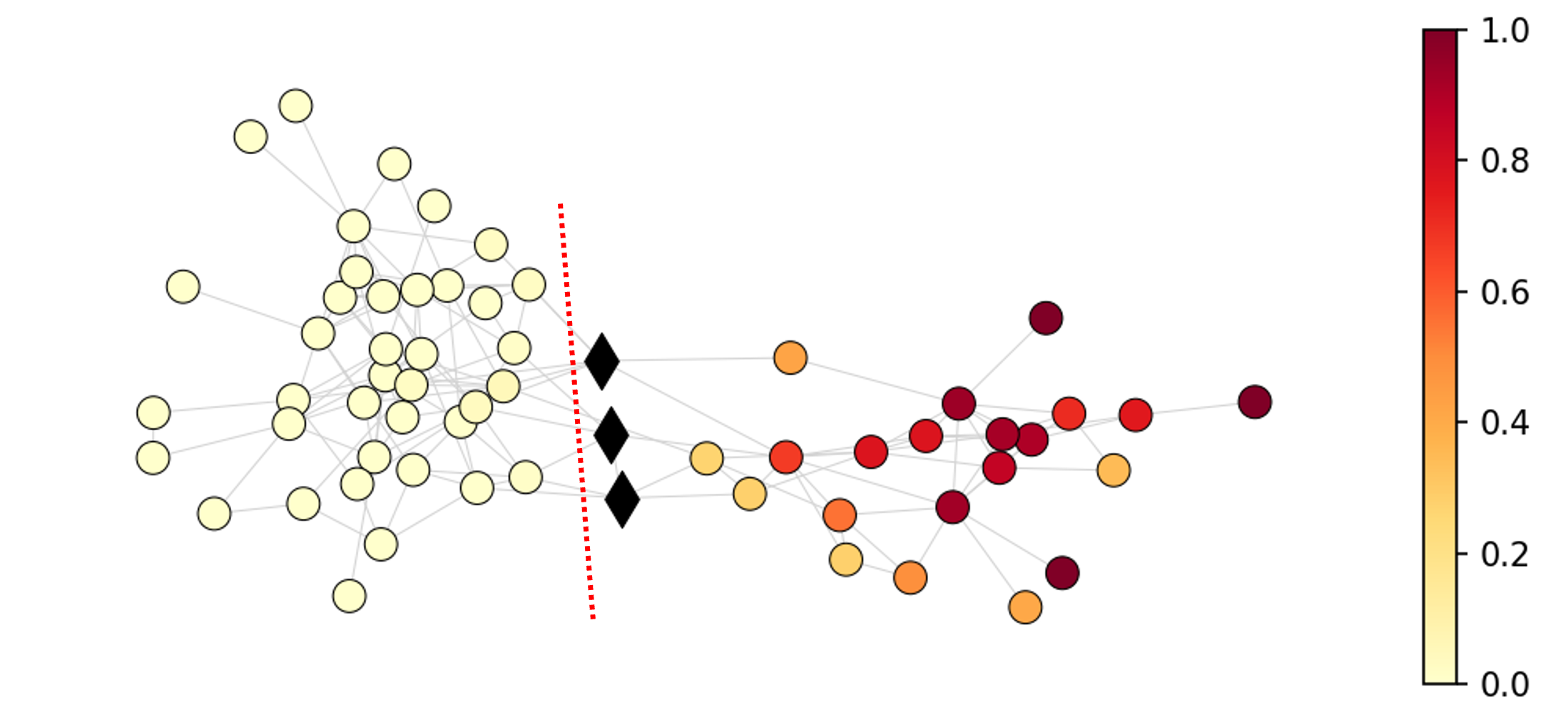}
    }
    \caption{Influence of the patching delay $T$ on which nodes to vaccinate. The color indicates the probability of infection.}\label{fig:delay}
    \vspace{-2mm}
\end{figure*}

In our earlier work~\cite{lee2019transient}, we establish a tighter upper bound for $\x(t)$, which works for any value of $t$ with an initial condition $\x(0)$. If $x_i(0) \in \{0,1\}$ for all $i$, we have 
\begin{equation}
  \x(t) \,\preceq\, \xxx(t) = f(\yyy(t)),  \label{ordering}
\end{equation}
where $f(y) \!=\! 1 \!-\! e^{-y}$, and $\yyy(t)$ is given by
\begin{align}
  \yyy(t)  =& -\log(1-\x(0)) \nonumber \\ 
  &+ \sum_{k=0}^{\infty} \frac{(\beta t)^{k+1}}{(k\!+\!1)!}\lt[\A\diag(\bm{1} \!-\! \x(0))\rt]^k \A \x(0). \label{upper-bound}
\end{align}
This upper bound allows us to predict the likelihood of each node being infected after \emph{any} time $t$ from the initial condition $\x(0)$. That is, even if the patching delay $T$ is moderate to possibly large, we can utilize this upper bound to predict the infection status of each node at $T$ so that we can identify which nodes can be successfully vaccinated without being infected before $T$.

\subsection{Significance of Patching Delay $T$}

We here demonstrate how different patching delays $T$ can change which nodes to patch or vaccinate to maximize the number of saved nodes from the vaccination. Assume that limited patches are available at $t \!=\! 0$. Figure~\ref{fig:delay} shows three different snapshots of an infection process with the infection rate $\beta = 0.05$ on a small network at $t \!=\! 0, 10, 15$, each of which depicts the probability of each node being infected at time $t$. Figure~\ref{fig:delay}(a) shows an initial state at $t \!=\! 0$, where three nodes are initially infected (indicated by red circles, for which $x_i(0) \!=\! 1$). If the patching delay is $T \!=\! 0$, i.e., the patching takes into effect immediately, then it is ideal to patch or vaccinate the nodes near the infection sources. However, as shown in Figure~\ref{fig:delay}(b), if the patching delay takes $T \!=\! 10$, then it will be better to vaccinate three nodes, indicated by the diamond symbol. This is because they have a small chance of getting infected during the patching delay $T$ while preventing infection from getting through the rest of the network, once they are successfully patched or immunized. If the patching is applied to the nodes near the infection sources, most of the patching will be wasted as they are more likely to be infected before the patching takes into effect. Similarly, if the patching delay takes $T \!=\! 15$, a bit farther nodes should be vaccinated as their risk of infection during $T$ is low while they can stop a further propagation of infection when they are vaccinated.  

As demonstrated in Figure~\ref{fig:delay}, for effective epidemic control, it is deemed crucial to identify the `boundary' separating the healthy nodes from the infected region and patch the boundary nodes. However, it is non-trivial to find such a boundary as it changes depending on the value of $T$ and its size also varies. Thus, we develop a mathematical framework to find a `patching' boundary that effectively isolates the healthy nodes from the infected regions while having a small size to cope with limited patching scenarios.

\section{Proposed Framework}\label{sec:framework}
\begin{figure*}[t!]
    \centering
    \includegraphics[width=\textwidth]{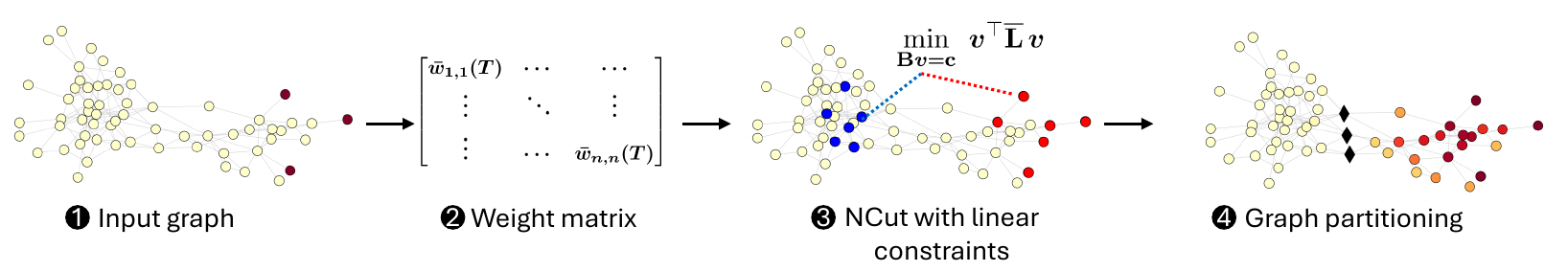}
    \vspace{-5mm}
    \caption{An overview of our proposed framework for effective delayed patching.}
    \label{fig:framework}
    \vspace{-4mm}
\end{figure*}

In this section, we provide the details of our proposed framework for effective delayed patching. It has four major steps as follows: constructing a weight matrix (Section~\ref{sec:weight}); formulating a graph partitioning problem to identify the patching boundary (Section~\ref{sec:con_ncut}); solving the problem (Section~\ref{sec:solution}); and using its solution to determine which nodes to patch or vaccinate under the budget constraint (Section~\ref{sec:patching}). Figure~\ref{fig:framework} illustrates an overview of the framework.

\subsection{Critical Edges and Edge Weights}\label{sec:weight}

To effectively identify the boundary separating the healthy nodes from the infected region, we below introduce a notion of `critical edges.' We then define the weight of an edge such that critical edges have higher weights, while non-critical ones have lower weights. 

For a particular patching delay $T$, there are four states that a node pair $(i,j) \in E$ forming an edge can take based on the infection status of nodes $i$ and $j$ at $T$, and they are categorized as the following two groups:
\begin{equation*}
(i,j) = 
\begin{cases}
C, & \text{if } (X_i(T), X_j(T)) = (0,1) \text{ or } (1,0) \\
C', & \text{if } (X_i(T), X_j(T)) = (0,0) \text{ or } (1,1)
\end{cases}
\end{equation*}
As the first group, C-type states correspond to edges that are susceptible to infection, which we label as \textit{critical edges}. In other words, a critical edge is an edge between a healthy node and an infected node (or vice versa). In contrast, $C'$-type states correspond to the non-critical edges that do not cause malware propagation, as infection cannot spread over healthy-healthy or infected-infected edges. Next, with the predicted infection probability $\hat{\x}(T)$ from (\ref{upper-bound}), we define the weight of an edge between nodes $i$ and $j$ as:
\begin{equation}\label{wT}
    w_{i,j}(T) = a_{ij} \left[ \hat{x}_i(T) ( 1 \!-\! \hat{x}_j(T) ) + ( 1 \!-\! \hat{x}_i(T)) \hat{x}_j(T) \right].
\end{equation}
Note that this edge weight $w_{i,j}(T)$ will encode our predicted state of the infection after time $T$ and can be interpreted as the probability of edge $(i,j)$ being critical at $T$. The weight value will get larger along the infection boundary, but it becomes smaller for edges well inside either susceptible or infected region, where $\hat{x}_i(T)$ and $\hat{x}_j(T)$ are both close to $0$ or $1$.

\subsection{Partitioning with Normalized Cut}\label{sec:con_ncut}

The problem is then to identify the boundary along the critical edges after the patching delay $T$ (to which a limited number of patches are applied), which can separate the most likely infection region from the nodes that would still be healthy by time $T$. To this end, we start with formulating the problem as a graph partitioning problem on our weighted graph. While several graph partitioning problems have been introduced in the literature, we here adopt the problem formulation of normalized cut (NCut), which is rooted in the spectral graph theory~\cite{shi2000normalized, jia2014latest, von2007tutorial}.

Consider a connected, weighted graph $G$, where each edge $(i,j) \in E$ is associated with a positive weight $w_{ij}$. To partition this graph (or the node set $N$) into two partitions $U$ and $U^c$, we find a cut between the two partitions, where the size of the cut is defined as
\begin{equation*}
\text{Cut}(U,U^c) \triangleq \sum_{i\in U}  \sum_{j\in U^c} w_{ij},
\end{equation*}
i.e., the sum of the weights of edges crossing the boundary between $U$ and $U^c$. The set of edges contributing to $\text{Cut}(U,U^c)$ is called a cut-set. The NCut problem is to solve the following:
\begin{equation}
    \min_{U \subset N} \text{NCut}(U) = \min_{U \subset N} \left( \frac{\text{Cut}(U, U^c)}{\Vol(U)} + \frac{\text{Cut}(U, U^c)}{\Vol(U^c)} \right)
    \label{ncut}
\end{equation}
where $\Vol(U)$ and $\Vol(U^c)$ are the sum of edge weights in $U$ and $U^c$, respectively. In other words, the NCut problem aims to find the minimum cut between the two partitions (clusters) while balancing the sum of edge weights in each partition. It is known that the NCut problem leads to a more balanced minimum cut than the pure minimum cut, which favors cutting small sets of isolated nodes in the graph~\cite{shi2000normalized}.

To provide a clean interface between our weighted graph setting and the NCut problem (which partitions along the edges with the minimum
weights), we “flip” the edge weights as
\begin{align}
\overline{w}_{i,j}(T) &\triangleq a_{ij} - w_{i,j}(T) \nonumber \\
&= a_{ij} \left[ \hat{x}_i(T) \hat{x}_j(T) + (1 \!-\! \hat{x}_i(T)) (1 \!-\! \hat{x}_j(T)) \right],\label{wT'}
\end{align}
where $w_{i,j}(T)$ is from (\ref{wT}). This way, the critical edges will now have the lowest weights, and all the logic so far will carry over to this ‘flipped’ version of edge weights.

Since the NCut problem in (\ref{ncut}) is NP-hard, we adopt its relaxed problem by relaxing the integer-valued solution, which takes $1$ or $-1$ for each node to indicate $U$ or $U^c$ it belongs to, to take arbitrary real values~\cite{shi2000normalized, jia2014latest, von2007tutorial}. To proceed, we define $d_i \!=\! \sum_j \overline{w}_{i,j}(T)$ to be the `generalized' degree of node $i$. Let $\D \!=\! \diag(d_1,d_2,\ldots, d_n)$ be the degree matrix, and let $\W \!=\! \lt[ \overline{w}_{ij}(T)\rt]$ be the $n \times n$ weight matrix, where edge weights $\overline{w}_{ij}(T)$ are given in (\ref{wT'}). The Laplacian and normalized Laplacian matrices of our weighted graph $G$ are defined as $\LL \!=\! \D - \W$ and $\overline{\LL} = \D^{-1/2} \LL \D^{-1/2}$, respectively. Then, the (relaxed) NCut problem is given by
\begin{equation}\label{norcut}
\begin{aligned}
    \min_{\vv \in \Rn^n} \vv^\top \overline{\LL} \vv & \\ 
    \text{subject to } \|\vv\|^2 =\Vol({N}) 
    & \text{ and } \vv^\top\D^{1/2}\bm{1} =0.
\end{aligned}
\end{equation}
This boils down to finding the eigenvector corresponding to the smallest non-zero eigenvalue of the normalized  Laplacian $\overline{\LL}$ and evaluating the ``sign" of each component of the eigenvector to partition the graph into two clusters.

In real-world scenarios, the number of patches available is limited and thus can be fewer than the number of (critical) edges in the cut-set that is identified as a solution to the NCut problem in (\ref{norcut}). While we could explicitly incorporate a constraint on the number of patches (or the budget constraint) into the NCut problem, there may be no solution to the problem. To address this potential issue, we impose the budget constraint separately after solving the NCut problem. In addition, we observe that a straightforward application of the solution to the NCut problem can lead to a cut partitioning either healthy or infected region inside, instead of separating one region from the other. See Figure~\ref{fig:partition} for an example. Thus, to steer the solution toward a more meaningful boundary of critical edges, we add linear constraints to the problem in (\ref{norcut}) in such a way that a small group of nodes whose status is definitively known to be healthy or infected at time $T$ is forced into their corresponding healthy or infected group, respectively. This constrained NCut problem remains tractable while producing a cut that clearly isolates high-risk nodes from the other ones.

\vspace{1mm}
\noindent \textbf{Choosing constrained nodes.} In what follows, we explain how to choose a small group of nodes for the linear constraints into the NCut problem. We first label nodes as `infected' or `healthy' using the predicted infection probability $\xxx(T)$ from (\ref{upper-bound}). Specifically, node $i$ is labeled as infected if $\hat{x}_i(T) > 0.5$ and healthy otherwise. From these labels, we select at most $10\%$ of nodes to be imposed as hard linear constraints for the NCut problem in (\ref{norcut}). The selection process is done as follows:

\begin{itemize}[itemsep=2pt,leftmargin=1.2em,topsep=2pt]
    \item \textit{Infected constraints:} We choose initially infected nodes and their one-hop neighbors as constrained nodes to be in the group of infected nodes. Since they are definitively infected or have the highest immediate risk of infection, they are used as `prompt' nodes to guide the partitioning of most likely infected nodes by the patching delay $T$.
    \item \textit{Healthy constraints:} For the nodes labeled healthy, we calculate the smallest shortest-path distance from any one of infection sources to each of them, sort them in a decreasing order of the shortest-path distance, and choose the top-$K$ nodes (having the longest shortest-path distance). They serve as `anchor' nodes in the healthy side of the partition.
\end{itemize}

\noindent In this work, the value of $K$ is chosen such that the total number of nodes in both constraints amounts to 10\% of the entire node set $N$ to steer the partition toward a meaningful boundary of critical edges without over-constraining the problem.

The selected constrained nodes are encoded as a set of linear constraints, which can be represented as 
\begin{equation*}
\B\vv \!=\! \cc,    
\end{equation*}
where each row of $\B\! \in\! \Rn^{m \times n}$ is a one-hot vector to indicate the location of each constrained node, and the corresponding entry in $\cc\! \in \!\Rn^m$ encodes its assigned label. Specifically, for each constrained node, we set $c_i\! =\! 1$ if it is labeled as infected and $c_i\! =\! -1$ if it is labeled as healthy. Note that the nodes that are not explicitly constrained do not appear in $\B$ or $\cc$. With these labeled constraints, the NCut problem in (\ref{norcut}) becomes
\begin{equation}
\label{eq:ncut_lincon}
\begin{aligned}
    \min_{\vv} \vv^\top \overline{\LL} \vv & \\ 
    \text{subject to } \|\vv\|^2 =\Vol({N}) 
    & \text{ and } \B\vv = \cc.
\end{aligned}
\end{equation}
Note that the orthogonality constraint $\vv^\top \D^{1/2}\bm{1} \!=\! 0$ is omitted. This is because any vector in the null space of $\overline{\LL}$ has all entries either equal to $+1$ or all equal to $-1$. However, by construction, our label constraints force at least one entry of $+1$ (infected) and another entry of $-1$ (healthy). Hence, no feasible $\vv$ can lie in the null space of $\overline{\LL}$, so we safely drop the orthogonality constraint.

\subsection{Solving Normalized Cut with Linear Constraints}\label{sec:solution}

To solve the constrained NCut problem in (\ref{eq:ncut_lincon}), we consider two methods, namely projected power method~\cite{xu2009fast} and augmented Lagrangian Uzawa method~\cite{o:00}. For the sake of completeness, we below review the principle of each method.

\vspace{1mm}
\noindent \textbf{Projected power method (PPM).} This method was designed to solve the following general problem:
\begin{equation*}
\begin{aligned}
   \max_{\vv} \vv^\top \M \vv  & \\ 
    \text{subject to } \|\vv\|^2 =\bm{1} 
    & \text{ and } \B\vv = \cc.
\end{aligned}
\end{equation*}
where $\M$ is a positive semidefinite matrix. In our setting, we set $\M\!=\!\alpha \I - \overline{\LL}$, where $\alpha$ is a sufficiently large value. $\Vol(N)$ is also a constant. Then, maximizing $\vv^\top\M\vv$ is equivalent to minimizing $\vv^\top \overline{\LL} \vv$ in (\ref{eq:ncut_lincon}).  We note that the feasible set is the intersection of the unit sphere and the hyperplane $\B\vv\!=\!\cc$. Thus, any feasible vector admits a decomposition $\vv\!=\!\nn_{0}+\zz,$ where $\nn_{0} \!=\! \B^\top(\B\B^\top)^{-1}\cc$ is the orthogonal projection of the origin onto $\B\vv \!=\! \cc$, and $\zz$ lies in the null space of $\B$ with a fixed norm $\|\zz\| \!=\! \gamma$, with $\gamma \!=\! \sqrt{1- \|\nn_{0}\|^{2}} > 0$. 
 
Starting from an initial vector $\vv_{0}$ that satisfies $\|\vv_{0}\| = 1$, we generate a sequence of iterates $\{\vv_{k}\}_{k\ge0}$ as follows:  At each step, we update the current iterate $\vv_{k}$ by first stretching it using the operator $\M$, then projecting the result back onto the hyperplane using the projection matrix $\p = \I - \B^\top(\B\B^\top)^{-1}\B$, and finally normalizing the projected vector to the length $\gamma$. Then, the new iterate $\vv_{k+1}$ is obtained by adding back $\nn_{0}$ after normalization. This procedure repeats until convergence and is summarized in Algorithm~\ref{ppm}. Here we assume that $\nn_{0}$ is not an eigenvector of $\M$ and has a nonzero component in the direction of the dominant eigenvector. Otherwise, a small perturbation can be added to satisfy this condition.

\setlength{\textfloatsep}{5pt}
\begin{algorithm}
\caption{Projected Power Method}
\begin{algorithmic}[1]
  \STATE $\p \leftarrow \I - \B^\top(\B \B^\top)^{-1}\B$
  \STATE $\nn_{0} \leftarrow \B^\top(\B \B^\top)^{-1}\cc$
  \STATE $\gamma \leftarrow \sqrt{1 - \|\nn_{0}\|^2}$
  \STATE $\vv \leftarrow \displaystyle\frac{\gamma \p \M \nn_{0}}{\|\p \M \nn_{0}\|} + \nn_{0}$
  \REPEAT
      \STATE $\zz \leftarrow \displaystyle \frac{\gamma\p\M\vv}{\|\p\M\vv\|}$
      \STATE $\vv \leftarrow \zz + \nn_{0}$
  \UNTIL{convergence}
  \STATE \textbf{return} $\vv$
\end{algorithmic}
\label{ppm}
\end{algorithm}

\vspace{1mm}
\noindent \textbf{Augmented Lagrangian Uzawa method (Uzawa)}. Although PPM can be used to solve the constrained NCut problem in (\ref{eq:ncut_lincon}) for most cases, it is inherently slow in practice due to its iterative process. Moreover, $\cc$ needs to be chosen carefully so that its resulting linear constraints, together with the volume constraint, do not lead to an ill-posed problem. To cope with these issues, we consider the following relaxed problem without having the volume constraint, as similarly done in~\cite{o:00} for image segmentation:
\begin{equation}
\label{eq:ncut_uzawa}
        \min_{\B\,\vv = \cc} \vv^\top \overline{\LL} \vv.
\end{equation}
This relaxation allows us to leverage the Uzawa method~\cite{uzawa:00}, which often leads to an accurate solution in a single iteration. 

\begin{figure*}[t!]
  \centering
  \subfigure[Constrained Nodes]{%
    \includegraphics[width=0.22\linewidth]{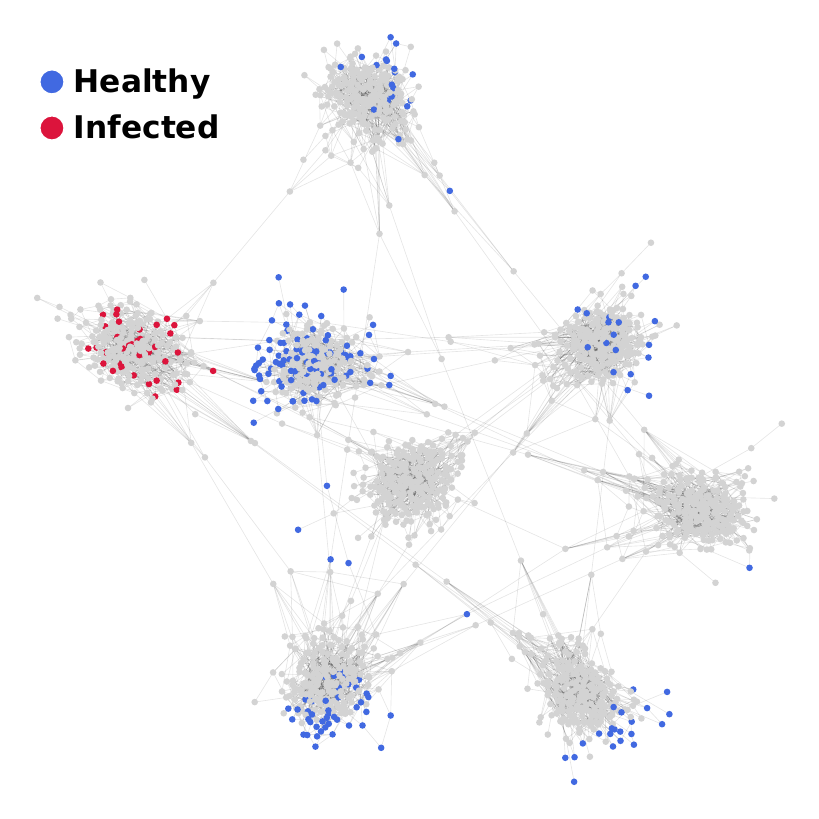}%
  }\hfill
  \subfigure[Vanilla NCut]{%
    \includegraphics[width=0.22\linewidth]{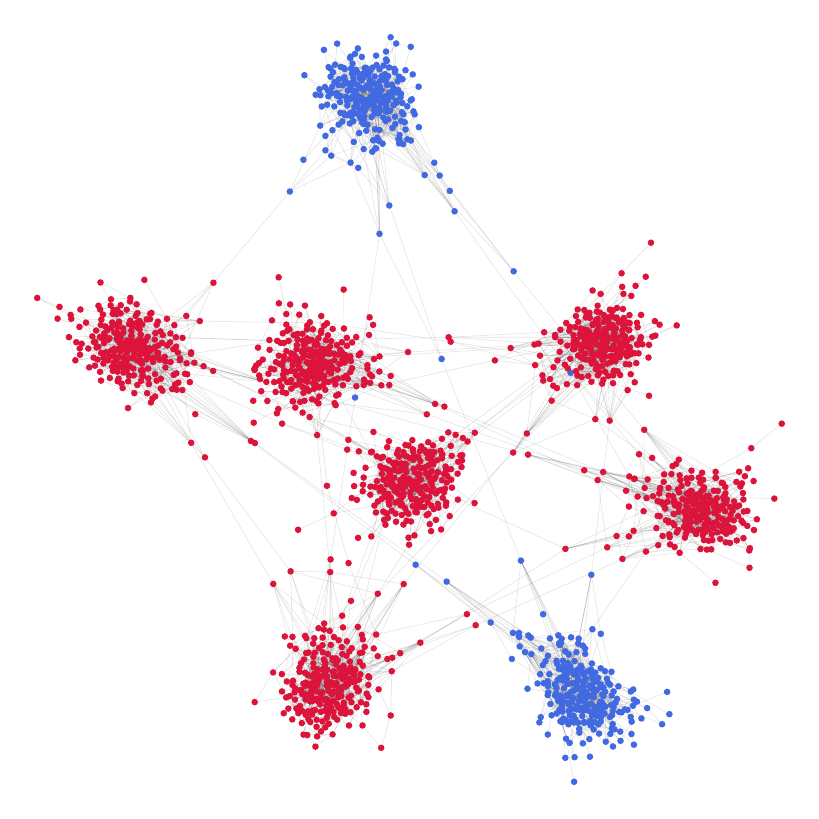}%
  }\hfill
  \subfigure[PPM]{%
    \includegraphics[width=0.22\linewidth]{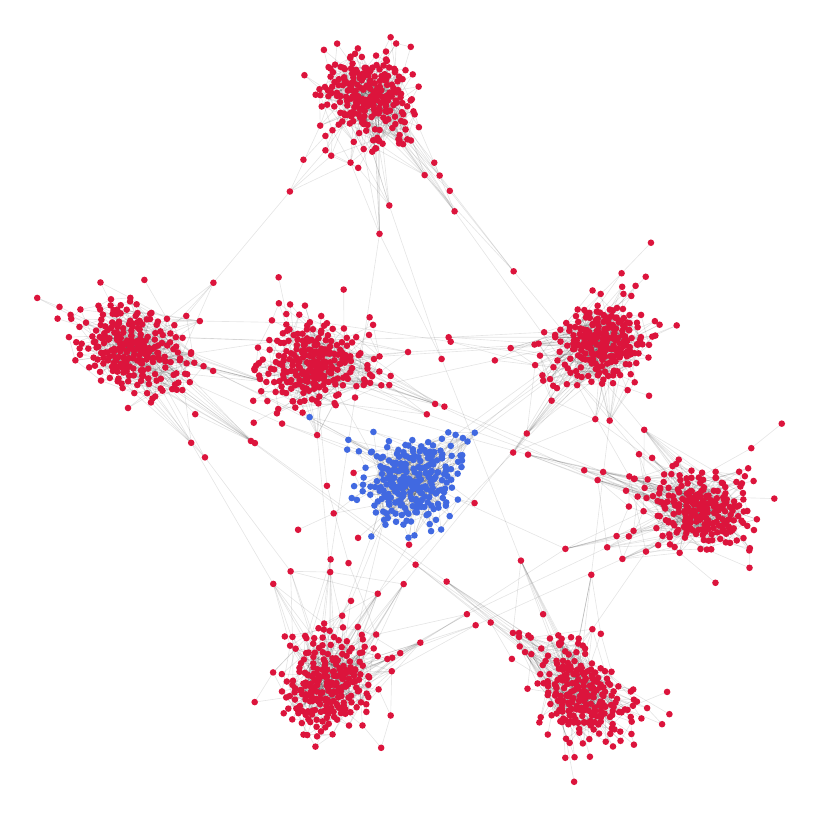}%
  }\hfill
  \subfigure[Uzawa]{%
    \includegraphics[width=0.22\linewidth]{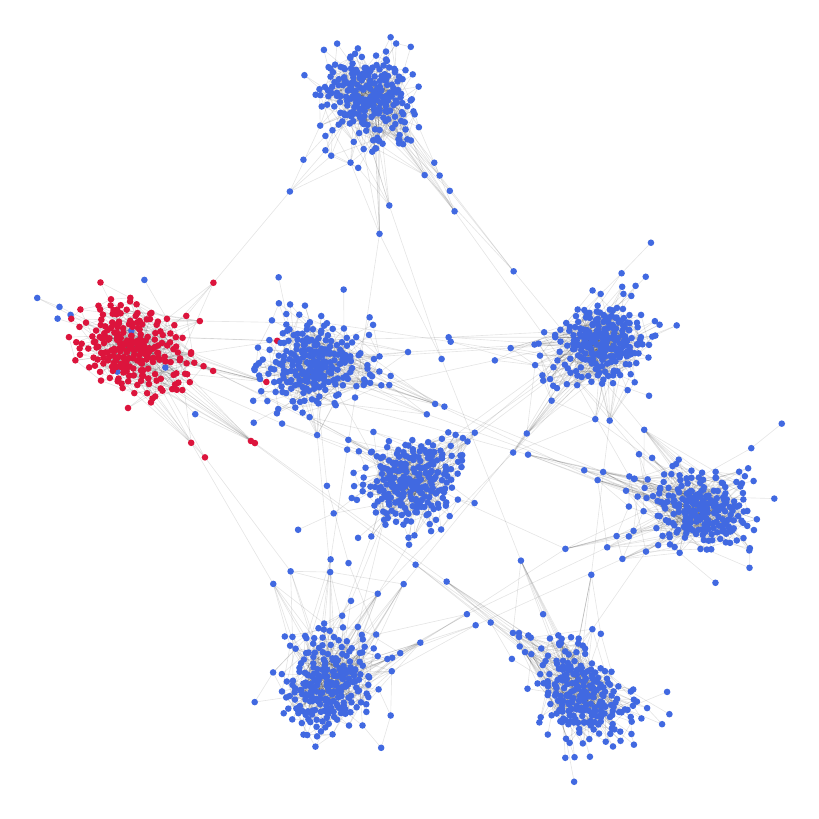}%
  }\hfill
  \subfigure[Runtime]{%
    \includegraphics[width=0.10\linewidth]{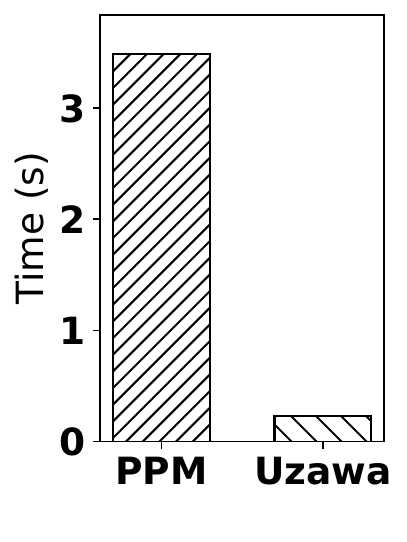}%
  }
  \vspace{-1mm}
  \caption{Performance comparison in partition quality and running time.}
  \label{fig:partition}
  \vspace{-5mm}
\end{figure*}
\vspace{1mm}

We begin by writing the Lagrangian function for the problem in (\ref{eq:ncut_uzawa}) as follows:
\begin{equation}
\label{eq:lag_func}
\begin{aligned}
        \mathcal{L}(\vv, \lambda)=\frac{1}{2}\vv^\top \overline{\LL}\vv  +\lambda (\B^\top \vv-\cc),
\end{aligned}
\end{equation}
where $\lambda$ is the Lagrange multiplier. The KKT conditions applied to this Lagrangian function yield
\begin{equation}\label{saddle}
\begin{pmatrix} 
\overline{\LL} & \B^\top \\ 
\B &  \bm{0} 
\end{pmatrix} 
\begin{pmatrix} 
\vv \\ 
\lambda 
\end{pmatrix} = \begin{pmatrix} 
0 \\ 
\cc
\end{pmatrix}.
\end{equation}
We can then use the Uzawa method to solve this indefinite system. While it has a very nice property that just one iteration provides a good approximate solution, we refer to \cite{uzawa:00, l:07, o:00} for mathematical details.

Specifically, by applying the Uzawa method to (\ref{saddle}), we can obtain $(\vv_1, \lambda_1)$ in the first iteration as follows: For given $(\vv_0, \lambda_0)$,
\begin{align*}
    (\overline{\LL}+\mu \B^\top\B) \vv_1 + \B \lambda_0 &= \mu \B^\top \cc \\ 
    \lambda_1 &= \lambda_0 + \mu (\B\vv_1-\cc),
\end{align*}
where $\mu$ is a sufficiently large value, i.e., $\mu \!\gg\! 1$. Setting $\lambda_0 \!=\! 0$ yields
\begin{equation}\label{yyy}
    (\overline{\LL}+\mu \B^\top \B) \vv_1 = \mu \B^\top\cc,
\end{equation}
with the error $\| \vv^* \!-\! \vv_1 \| \lesssim \sqrt{1/\mu}$, where $\vv^*$ is the solution to \eqref{saddle}. Thus, with $\mu \!\gg\! 1$, we can safely use $\vv_1$ in \eqref{yyy} as an approximate solution to \eqref{saddle}. In other words, we have the following solution to \eqref{saddle}:
\begin{equation}\label{approx}
    \vv \approx \mu  (\overline{\LL}+\mu \B^\top \B)^{-1} \B^\top\cc.
\end{equation}

We empirically observe that the Uzawa method yields a higher-quality partition than PPM and the vanilla NCut while being computationally faster than PPM. As an example, in Figure~\ref{fig:partition}, we show the partition quality of the vanilla NCut, PPM, and the Uzawa method, as well as the running times of PPM and the Uzawa method. The results were obtained on a synthetic graph of 2000 nodes, generated by the stochastic block model, with infection rate $\beta \!=\! 0.01$ and patching delay $T \!=\! 2$. As shown in Figure~\ref{fig:partition}, the Uzawa method identifies the infection region clearly and isolates it from the healthy nodes (the rest of the graph), while both NCut and PPM cut through the healthy region, leading to poorer-quality partitions. Furthermore, the Uzawa method runs roughly 30 times faster than PPM. Thus, in this work, we adopt the Uzawa method to solve our constrained NCut problem in (\ref{eq:ncut_lincon}).

\subsection{Node Selection for Patching under Budget Constraint}\label{sec:patching}

As a solution to the problem in (\ref{eq:ncut_lincon}), we obtain a cut-set, denoted as $E_c$, which is the set of `cross-boundary' edges between $U$ and $U^c$. What remains is how to choose a given number of nodes for patching from the cut-set $E_c$, especially when the budget constraint on the number of available patches (or vaccines) is smaller than the size of the cut-set, i.e., $|E_c|$. To this end, we propose the following greedy heuristic: We first find a healthy-side node with the highest degree (obtained from the original, unweighted adjacency matrix) in $E_c$ and then remove it, which in turn deletes all the edges associated with that node from $E_c$. We repeat this process on the reduced cut-set until it becomes empty or the budget is fully allocated. If the budget is still available, we consider all one-hop (yet not selected) neighbors of the nodes to be patched and repeat the process again until the budget is fully utilized.

The operation of the framework, as illustrated in Figure \ref{fig:framework}, can be summarized as follows. Given an input graph $G$, infection rate $\beta$, and patching delay $T$, we first calculate $\xxx(T)$ using \eqref{upper-bound} and construct a weight matrix $\W$ to identify a boundary along the critical edges. By patching or vaccinating the boundary nodes, we can effectively suppress the malware propagation from the infected region to healthy nodes. To find such a boundary, we formulate a graph partitioning problem as a relaxed NCut problem with linear constraints where a few nodes whose infection or healthy status is crystal clear are selected and placed in each group so that they can behave as anchor nodes to guide the partitioning process. We finally adopt the Uzawa method to solve the problem and identify which nodes to patch or vaccinate under the budget constraint.

\begin{figure*}[t!]
    \centering
    \subfigure[$T=15,\ n=1000 , \ k=3 $]{%
        \includegraphics[width=0.23\linewidth]{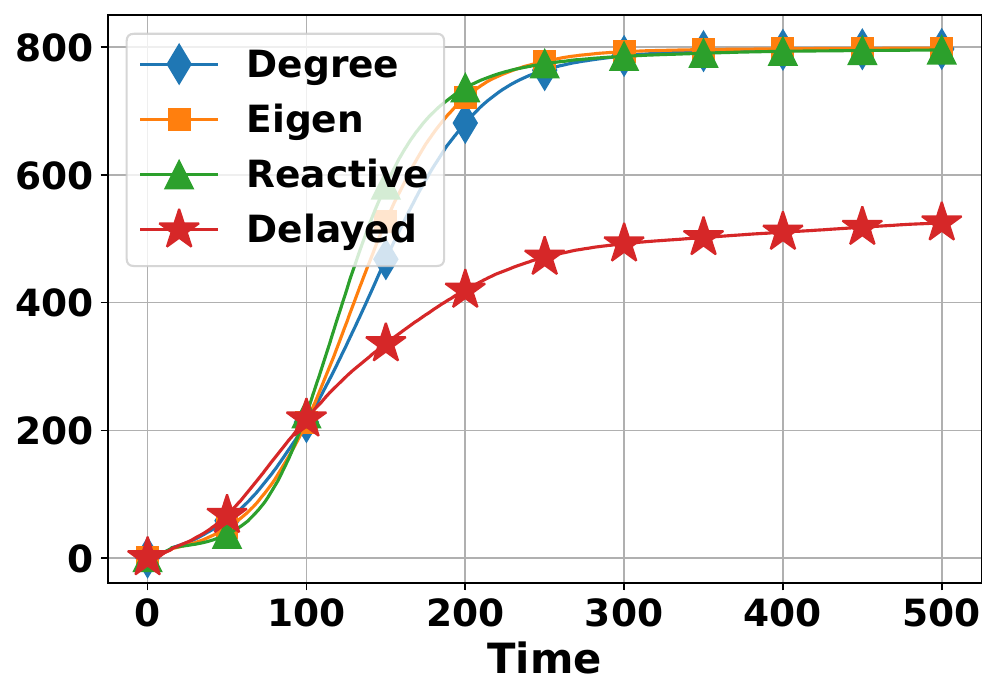}
    }
    \subfigure[$T=20,\ n=1000, \  k=3 $]{%
        \includegraphics[width=0.23\linewidth]{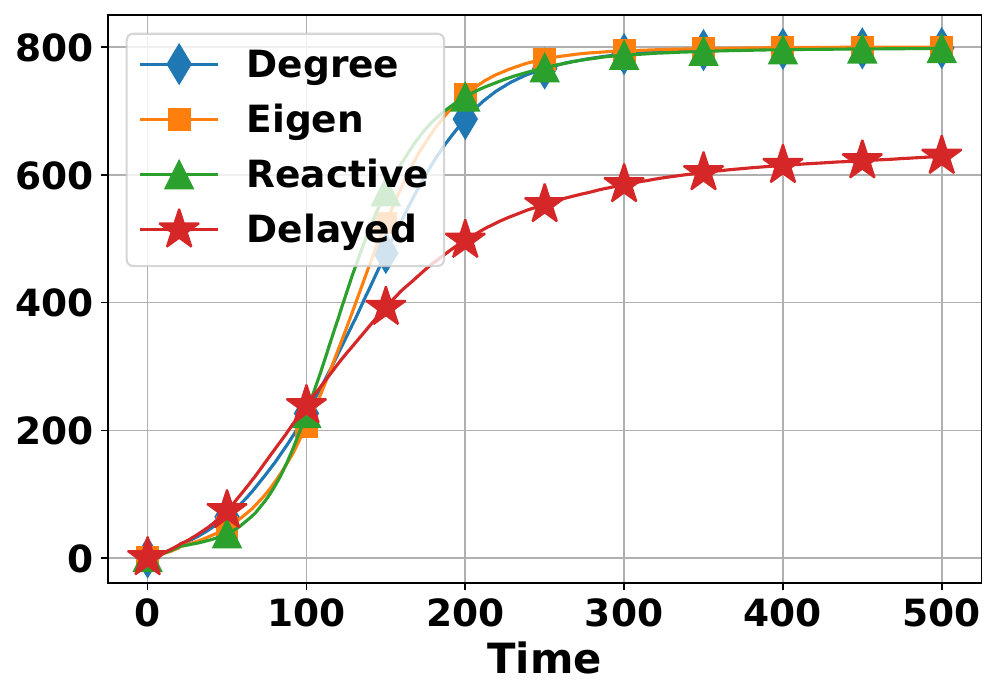}
    }
    \subfigure[$T=25,\  n=1000, \ k=3  $]{%
        \includegraphics[width=0.23\linewidth]{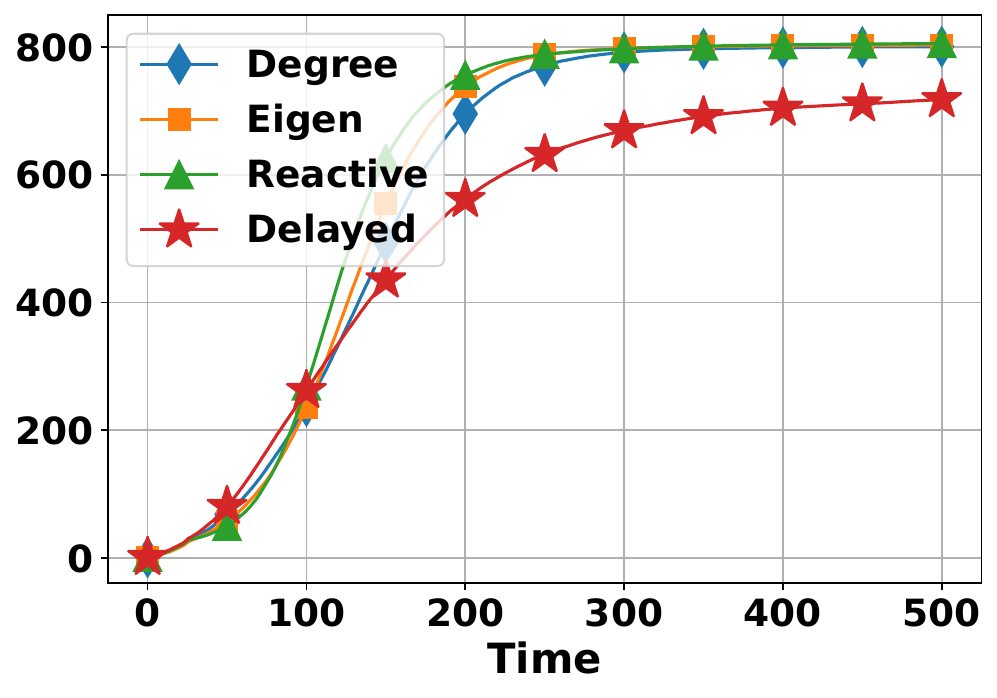}
    }
    \subfigure[$T=30,\ n=1000 , \ k=3 $]{%
        \includegraphics[width=0.23\linewidth]{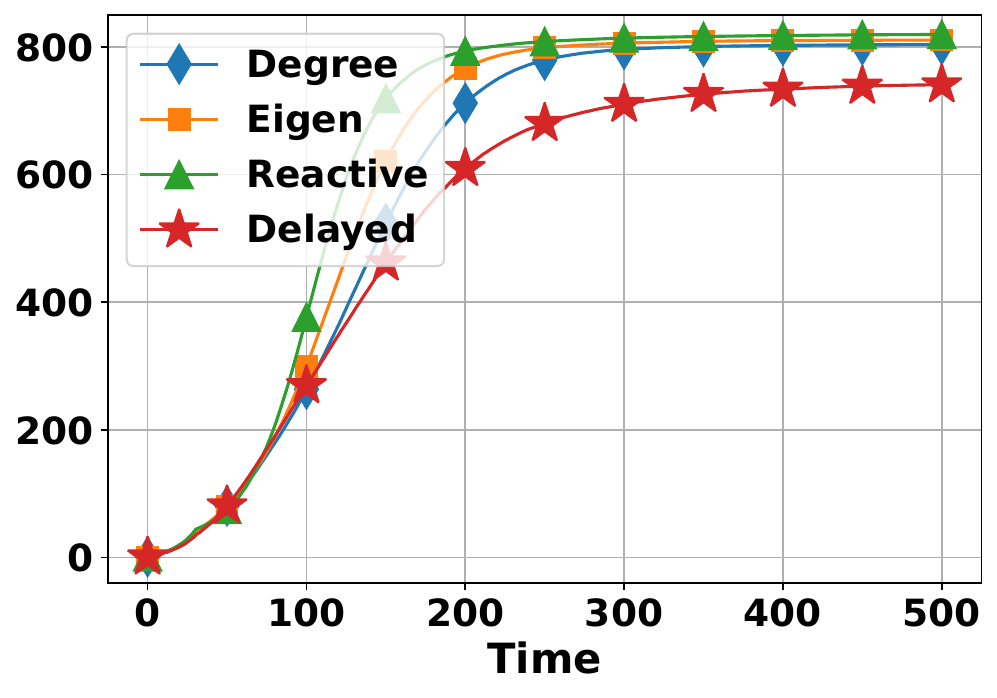}
    }
    \subfigure[$T=15,\ n=2000 , \ k=4 $]{%
        \includegraphics[width=0.23\linewidth]{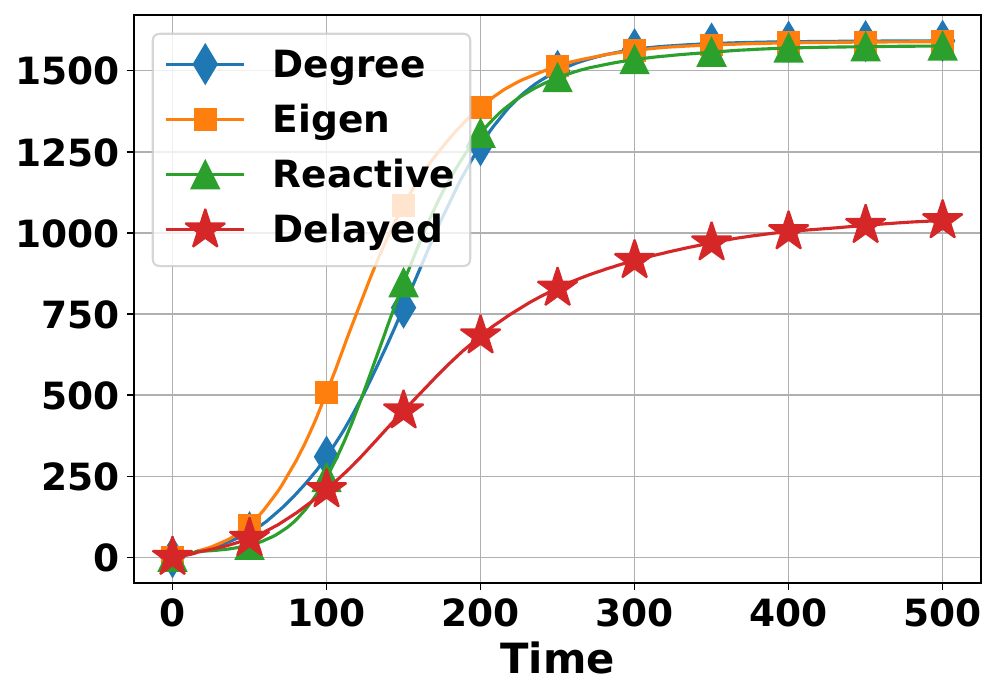}
    }
    \subfigure[$T=20,\ n=2000, \  k=4$]{%
        \includegraphics[width=0.23\linewidth]{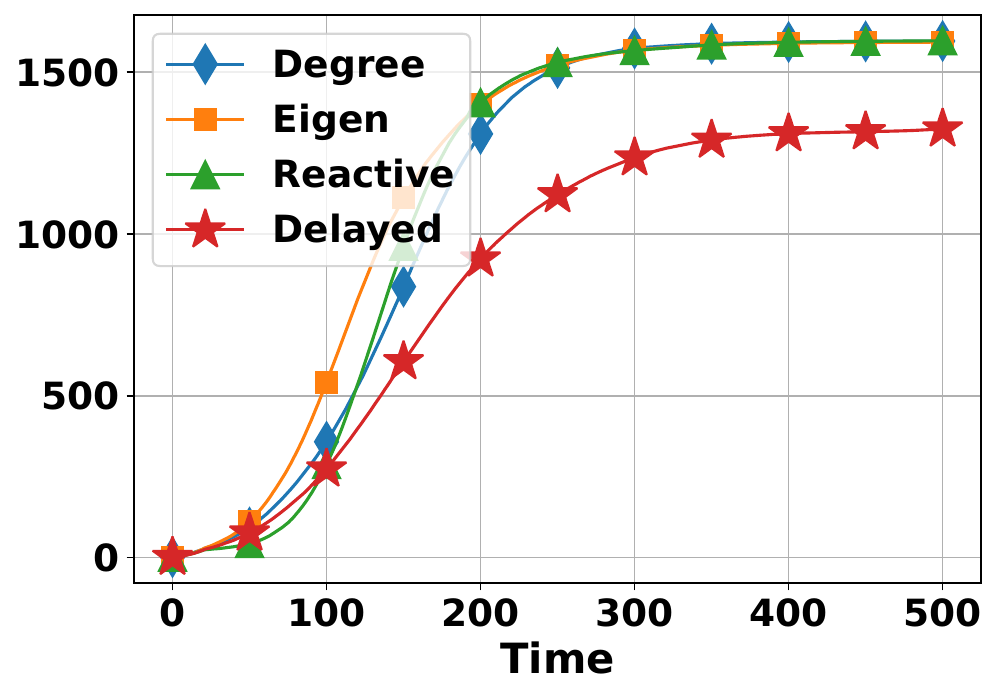}
    }
    \subfigure[$T=25,\ n=2000 , \ k=4 $]{%
        \includegraphics[width=0.23\linewidth]{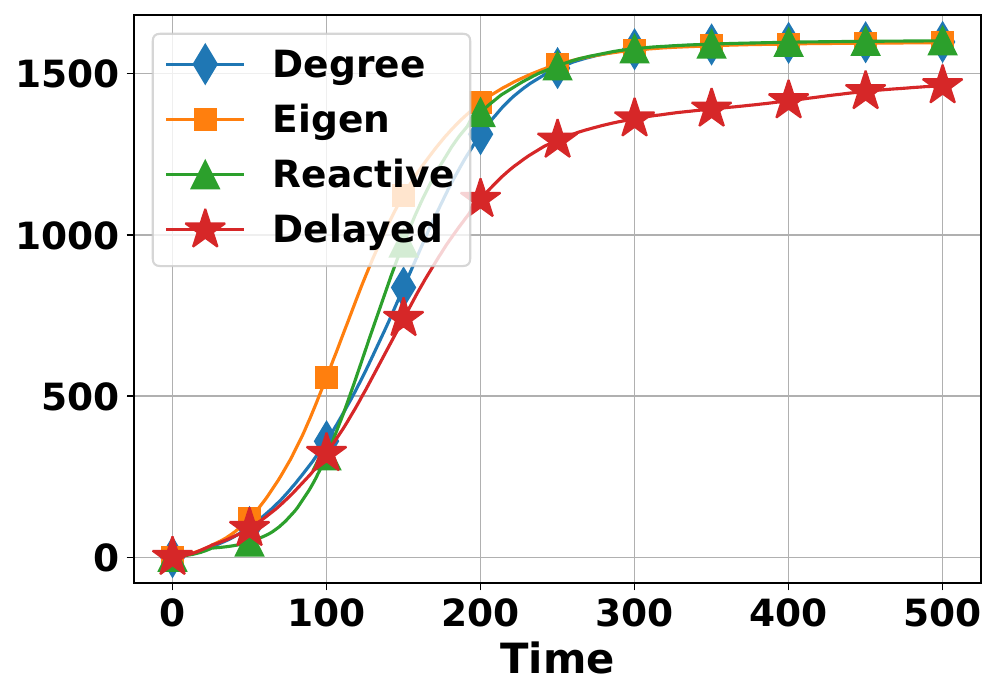}
    }
    \subfigure[$T=30,\ n=2000, \ k=4  $]{%
        \includegraphics[width=0.23\linewidth]{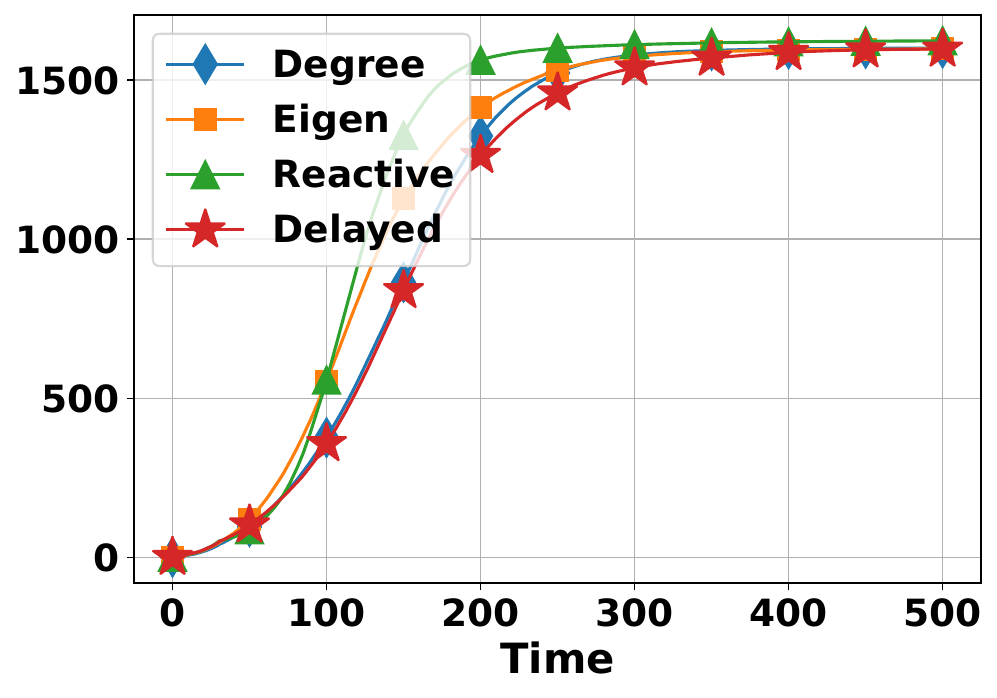}
    }
    
    \subfigure[$T=15,\ n=4000 , \ k=5 $]{%
        \includegraphics[width=0.23\linewidth]{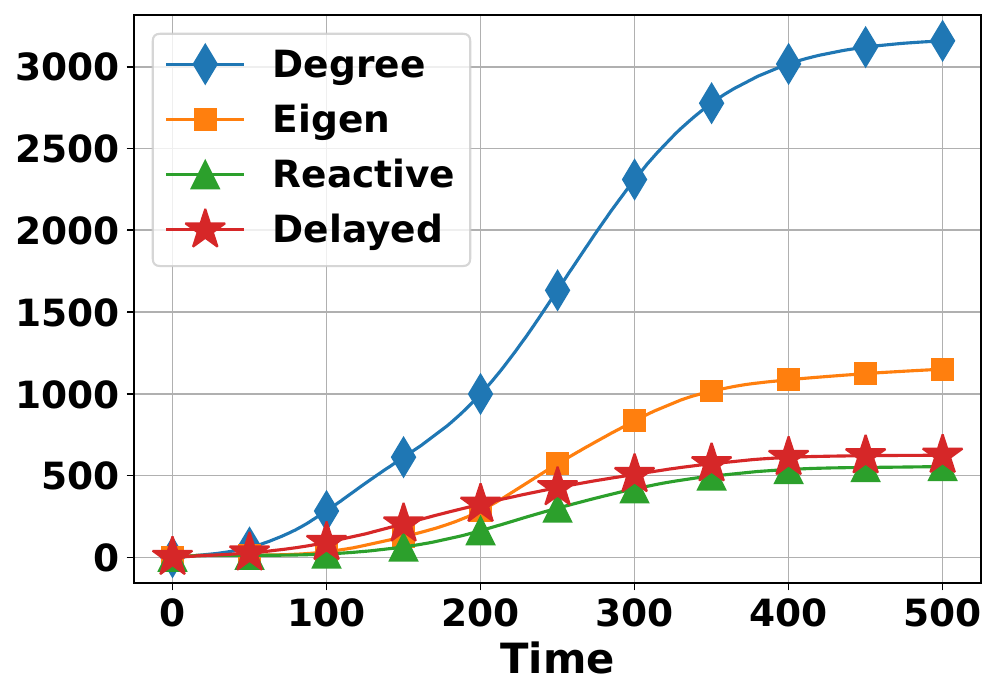}
        \label{fig:sbmk5t15}
    }
    \subfigure[$T=20,\ n=4000 , \ k=5 $]{%
        \includegraphics[width=0.23\linewidth]{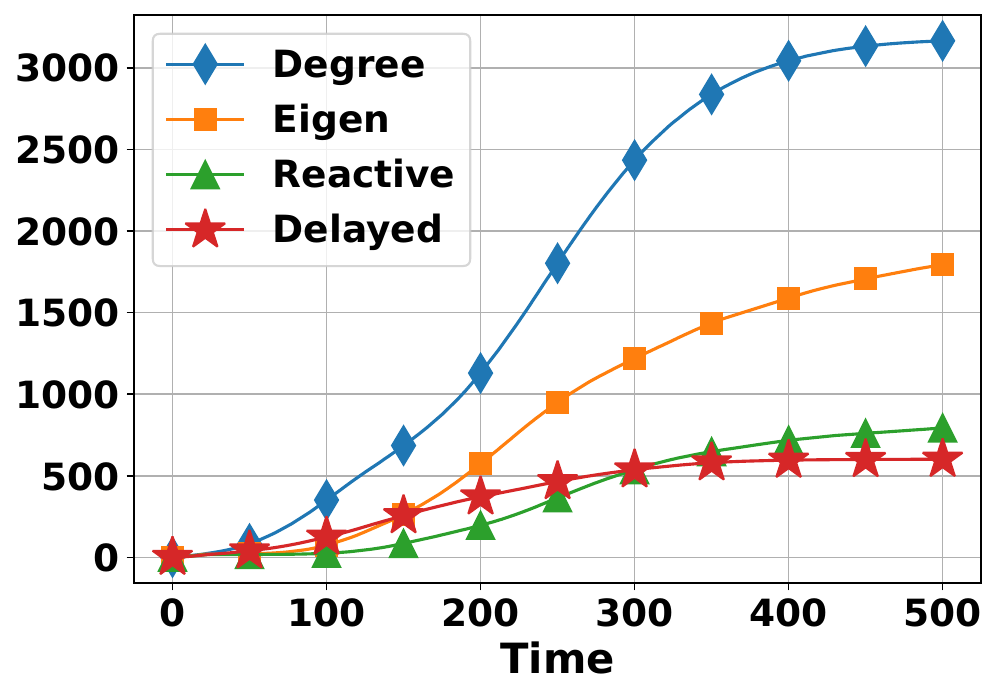}
        \label{fig:sbmk5t20}
    }
    \subfigure[$T=25,\ n=4000, \ k=5 $]{%
        \includegraphics[width=0.23\linewidth]{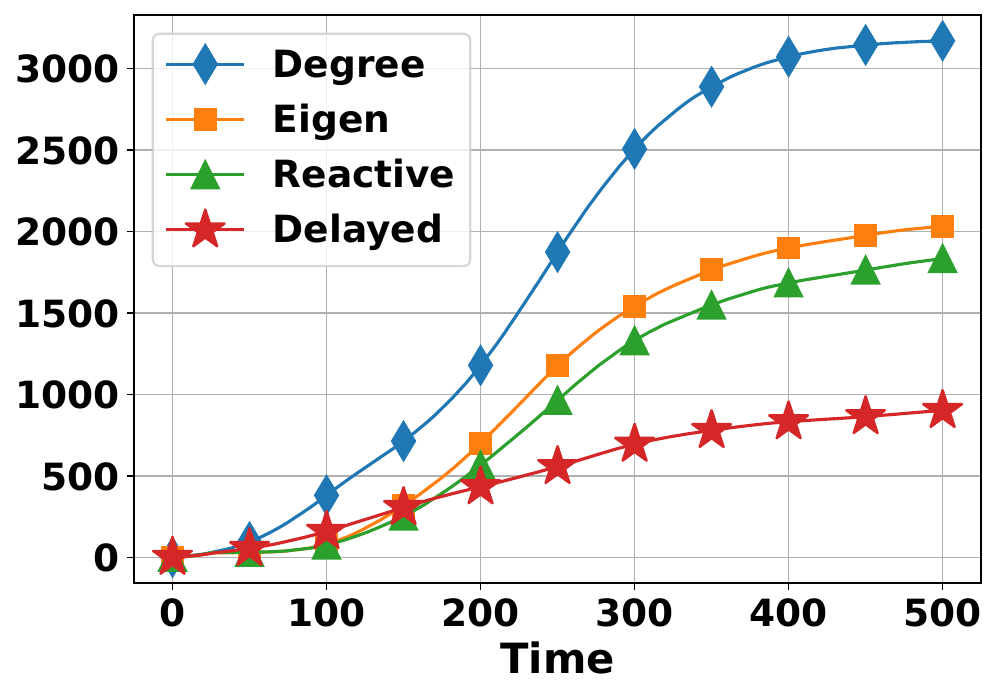}
        \label{fig:sbmk5t25}
    }
    \subfigure[$T=30,\ n=4000 , \ k=5  $]{%
        \includegraphics[width=0.23\linewidth]{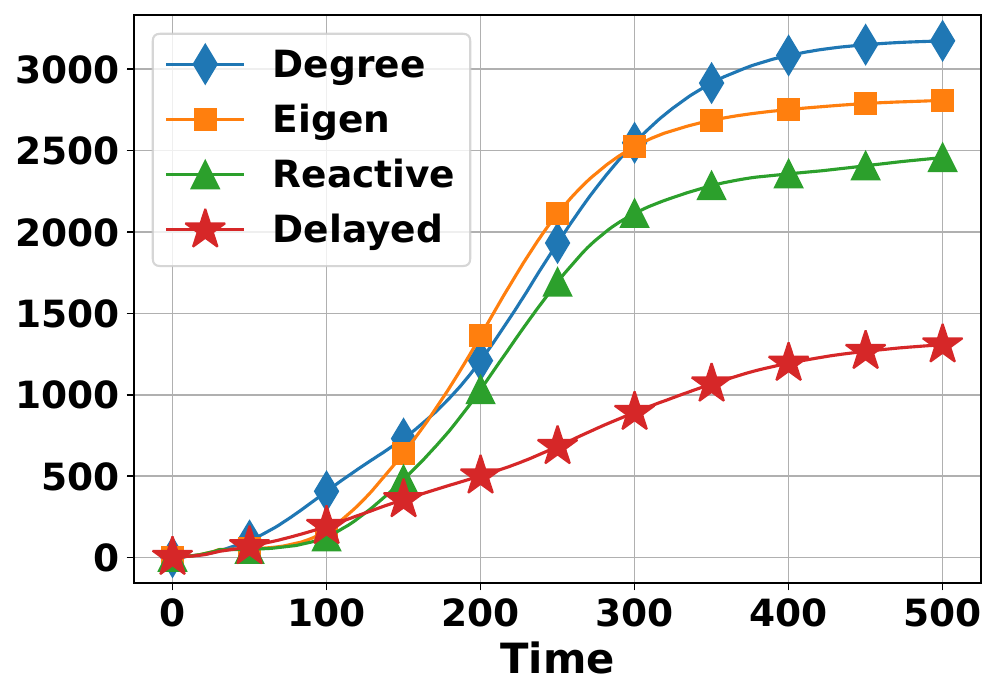}
        \label{fig:sbmk5t30}
    }

    \subfigure[$T=15,\ n=8000, \ k=6 $]{%
        \includegraphics[width=0.23\linewidth]{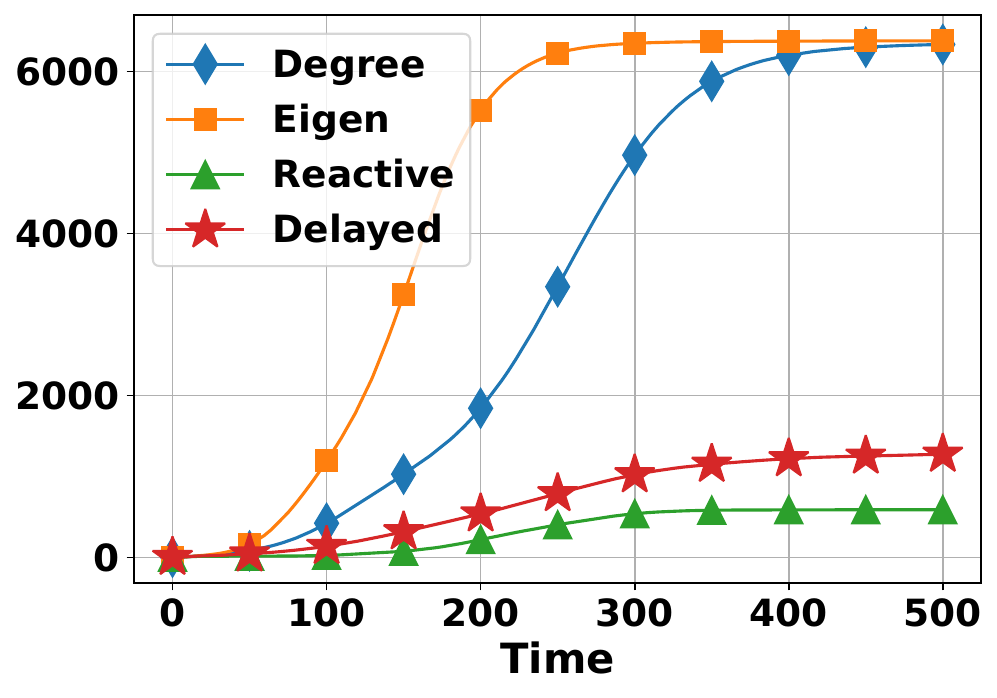}
        \label{fig:sbm_time_30000_7}
    }
    \subfigure[$T=20,\ n=8000 , \ k=6 $]{%
        \includegraphics[width=0.23\linewidth]{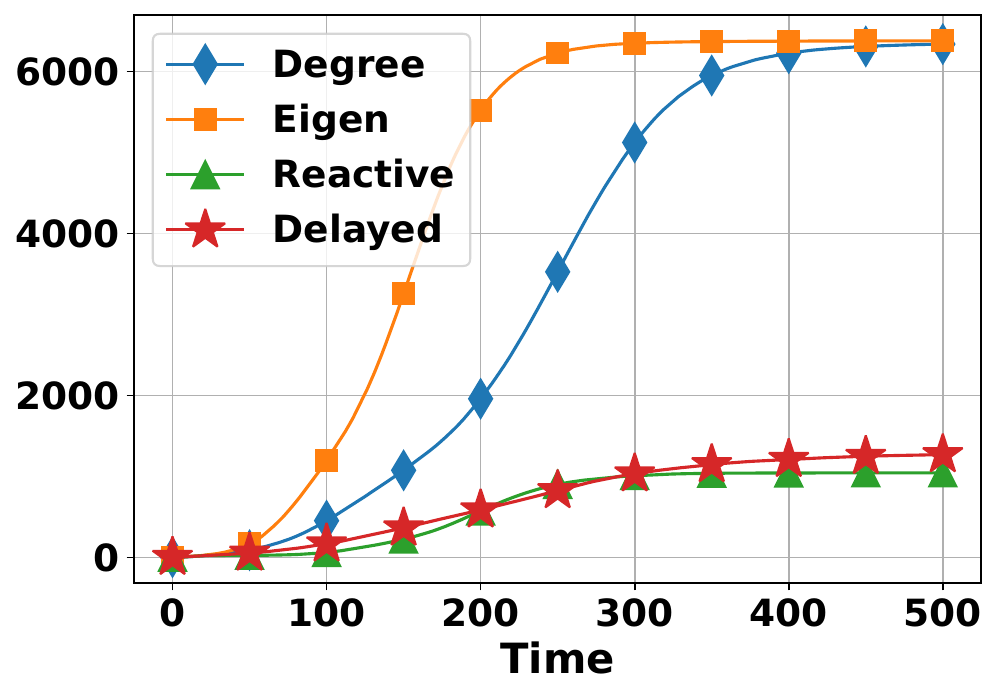}
        \label{fig:sbm_time_30000_8}
    }
    \subfigure[$T=25,\ n=8000 , \ k=6 $]{%
        \includegraphics[width=0.23\linewidth]{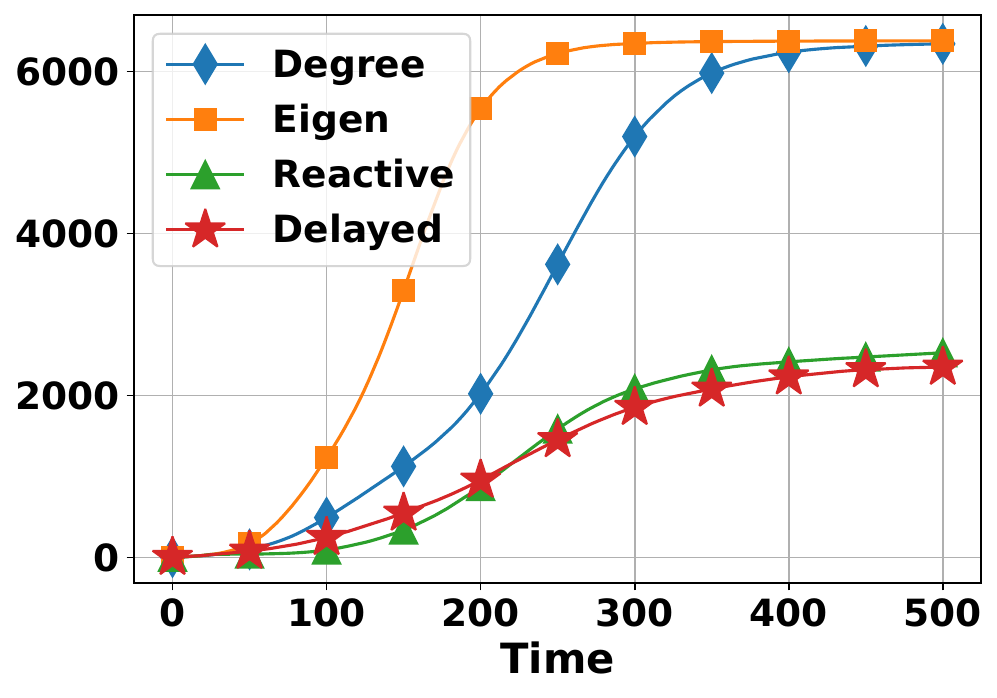}
        \label{fig:sbmk6t25}
    }
    \subfigure[$T=30,\  n=8000, \ k=6 $]{%
        \includegraphics[width=0.23\linewidth]{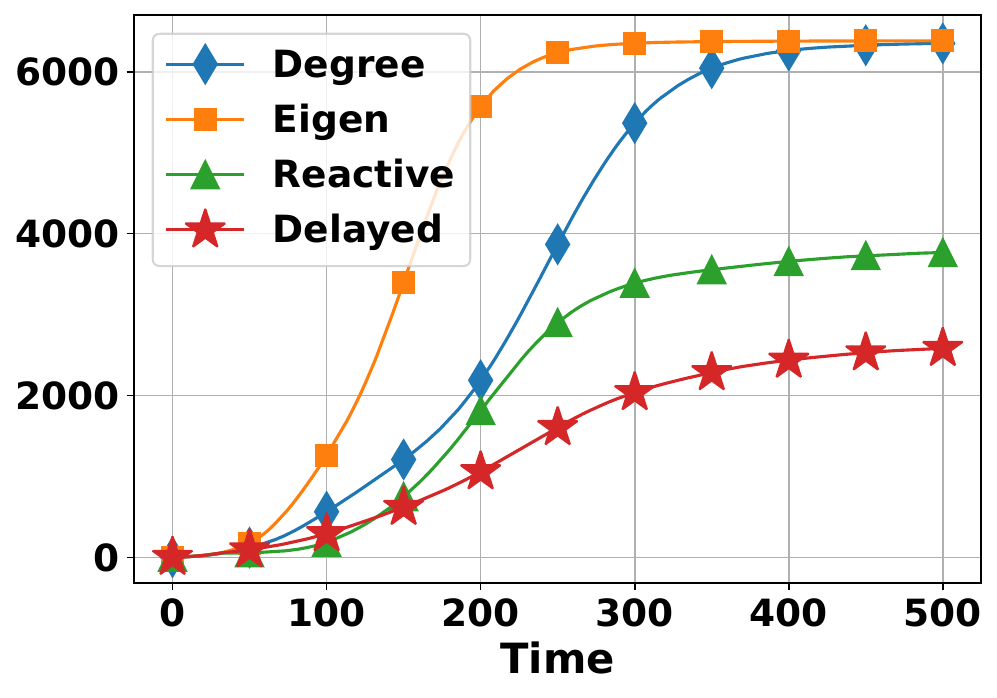}
        \label{fig:sbmk6t30}
    }
    \vspace{-2mm}
    \caption{The expected number of infected nodes by each vaccination policy on synthetic graphs.}
    \label{fig:sbm}
    \vspace{-5mm}
\end{figure*}

\section{Performance Evaluation}\label{sec:simulation}
\begin{figure*}[htbp]
  \centering
  \subfigure[$k=3$]{%
    \includegraphics[width=\textwidth]{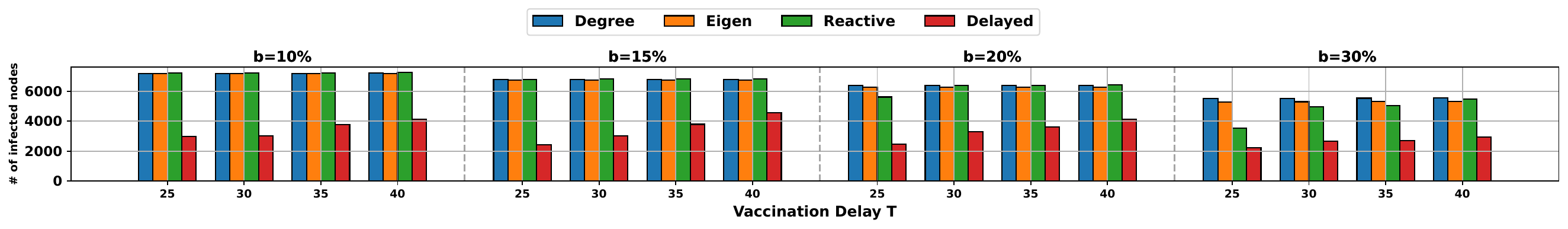}
  }\\[1ex]
  \vspace{-1.5em}
  \subfigure[$k=4$]{%
    \includegraphics[width=\textwidth]{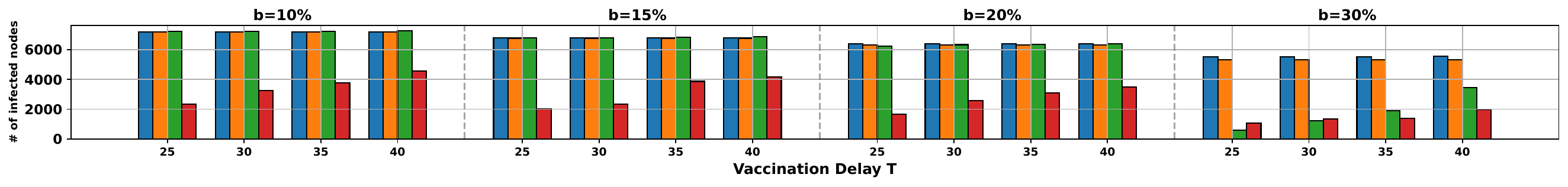}
  }\\[1ex]
  \vspace{-1.5em}
  \subfigure[$k=5$]{%
    \includegraphics[width=\textwidth]{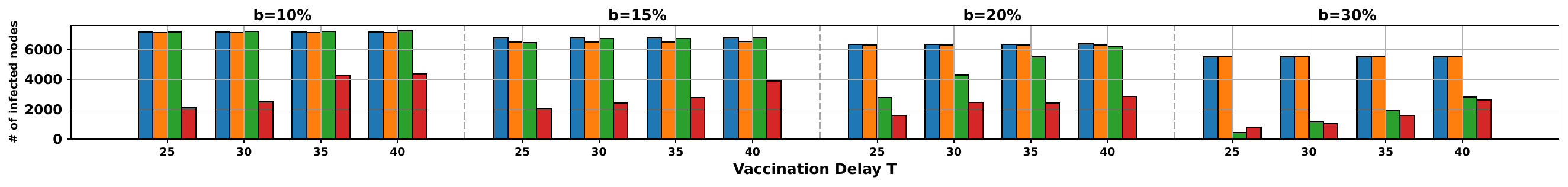}
  }\\[1ex]
  \vspace{-1.5em}
  \subfigure[$k=6$]{%
    \includegraphics[width=\textwidth]{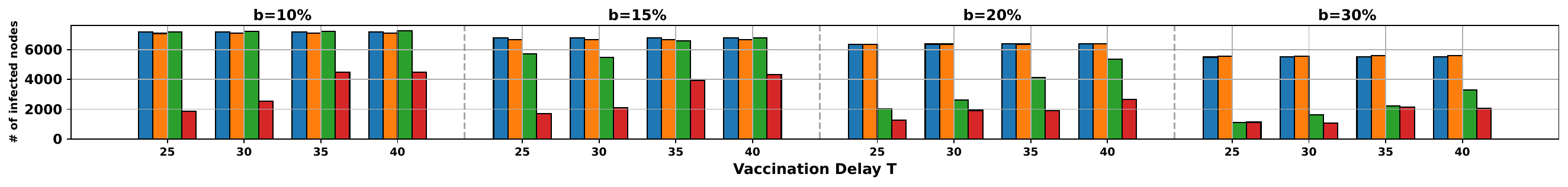}
  }
    \vspace{-3mm}
    \caption{The expected number of infected nodes at $t \!=\! 1000$ on a synthetic graph with $n \!=\! 8000$ when changing the values of $k$, $b$, and $T$.}
    \label{fig:num_infected_sbm}
    \vspace{-3mm}
\end{figure*}

\begin{figure*}[t!]
  \centering
  \subfigure[$T=5,\ b=10\%$]{%
    \includegraphics[width=0.25\linewidth]{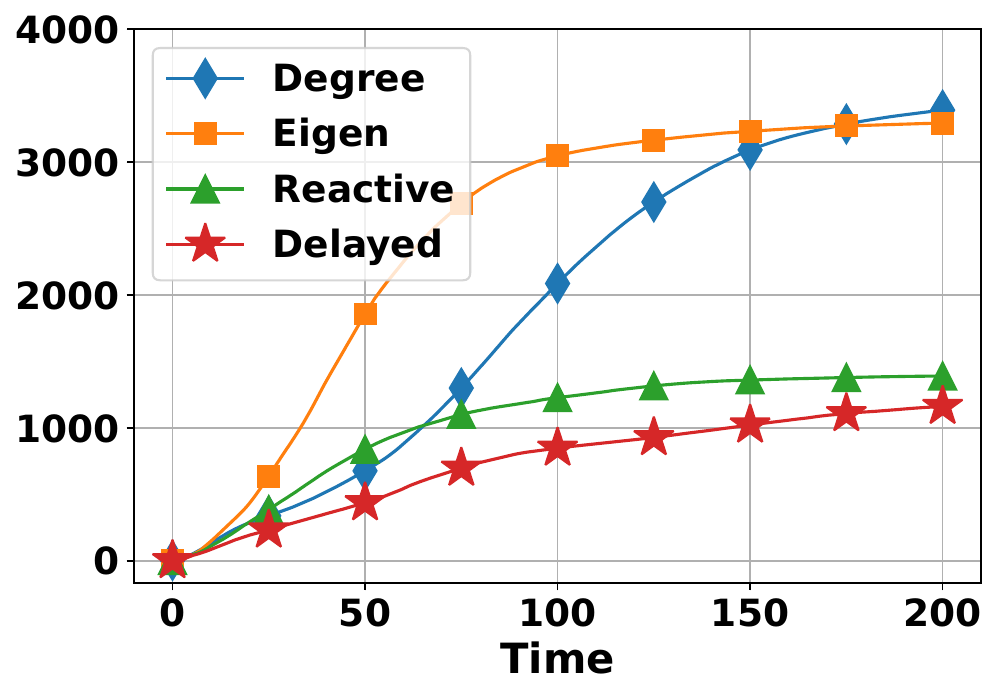}%
  }%
  \subfigure[$T=10,\ b=10\%$]{%
    \includegraphics[width=0.25\linewidth]{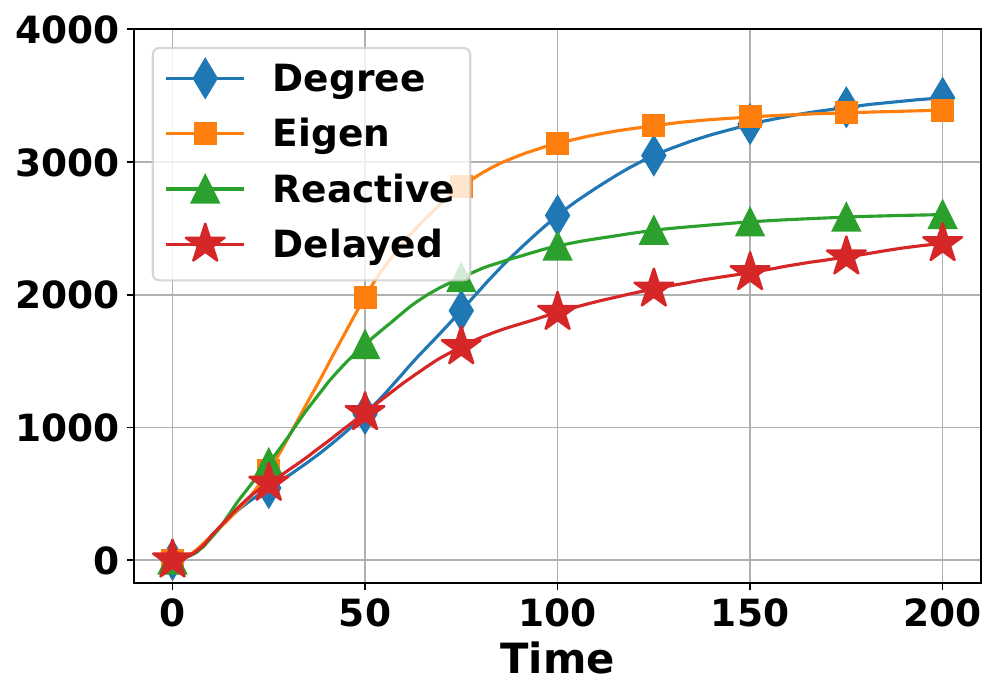}%
  }%
  \subfigure[$T=15,\ b=10\%$]{%
    \includegraphics[width=0.25\linewidth]{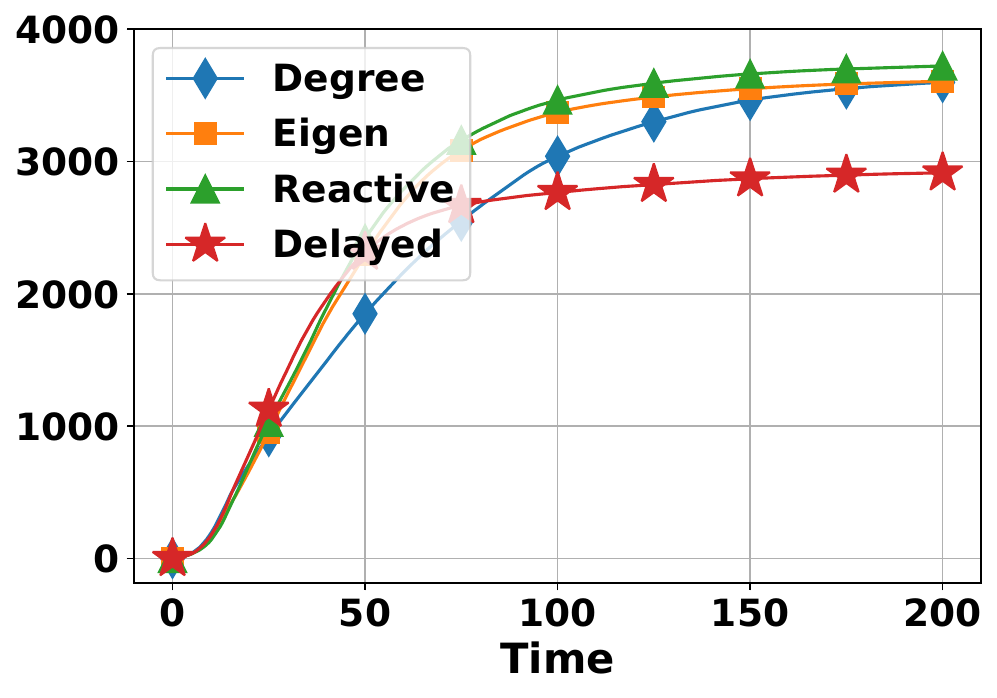}%
  }%
  \subfigure[$T=20,\ b=10\%$]{%
    \includegraphics[width=0.25\linewidth]{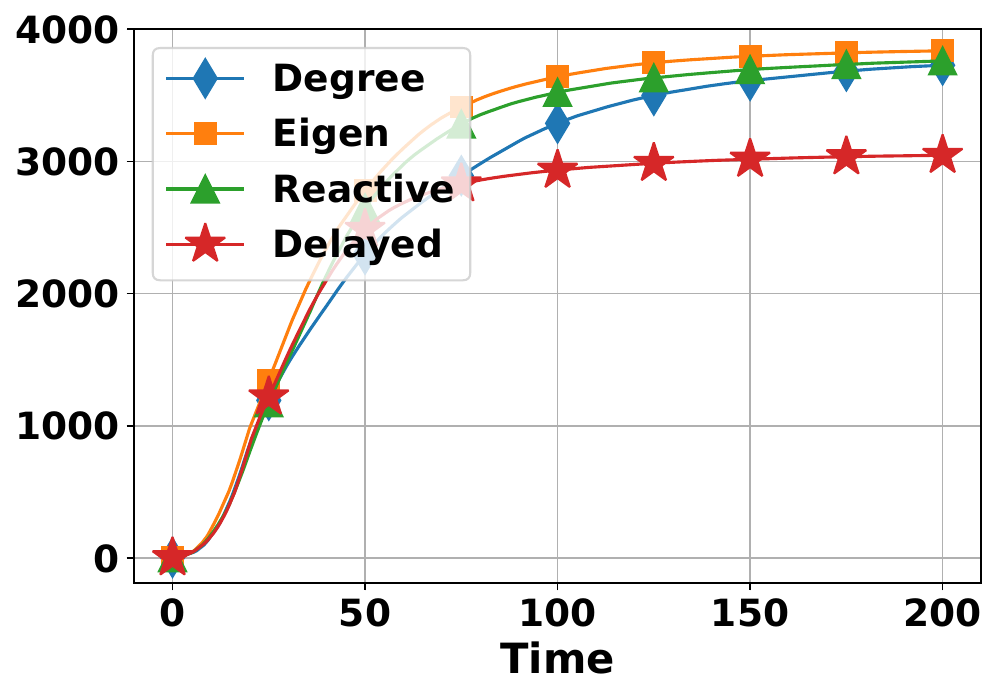}%
  }%
  \vspace{-1mm} 
  \subfigure[$T=5,\ b=20\%$]{%
    \includegraphics[width=0.25\linewidth]{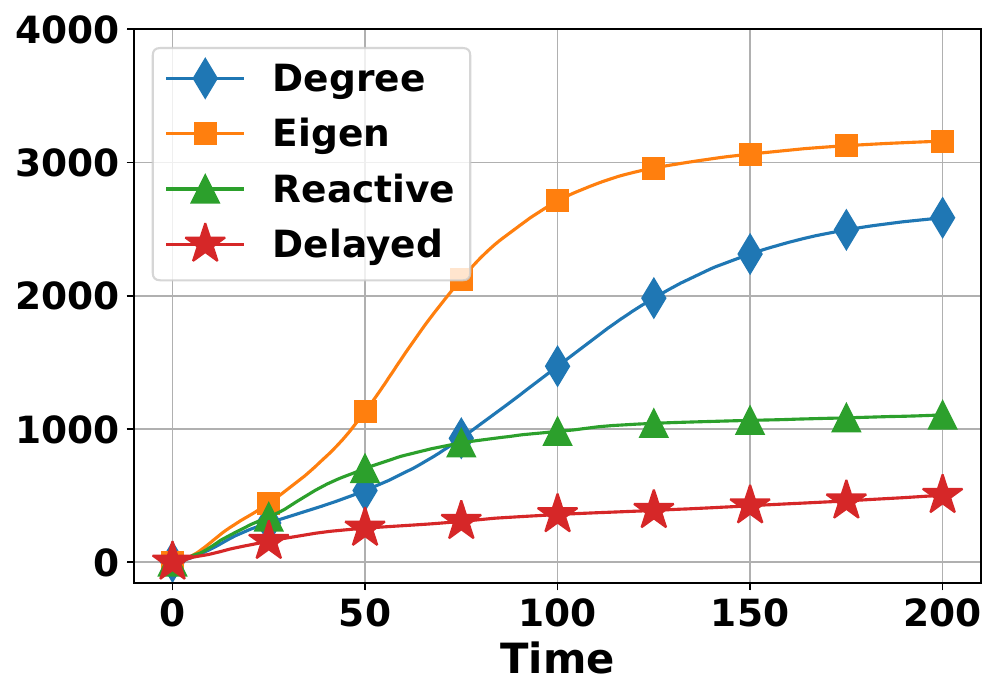}%
  }%
  \subfigure[$T=10,\ b=20\%$]{%
    \includegraphics[width=0.25\linewidth]{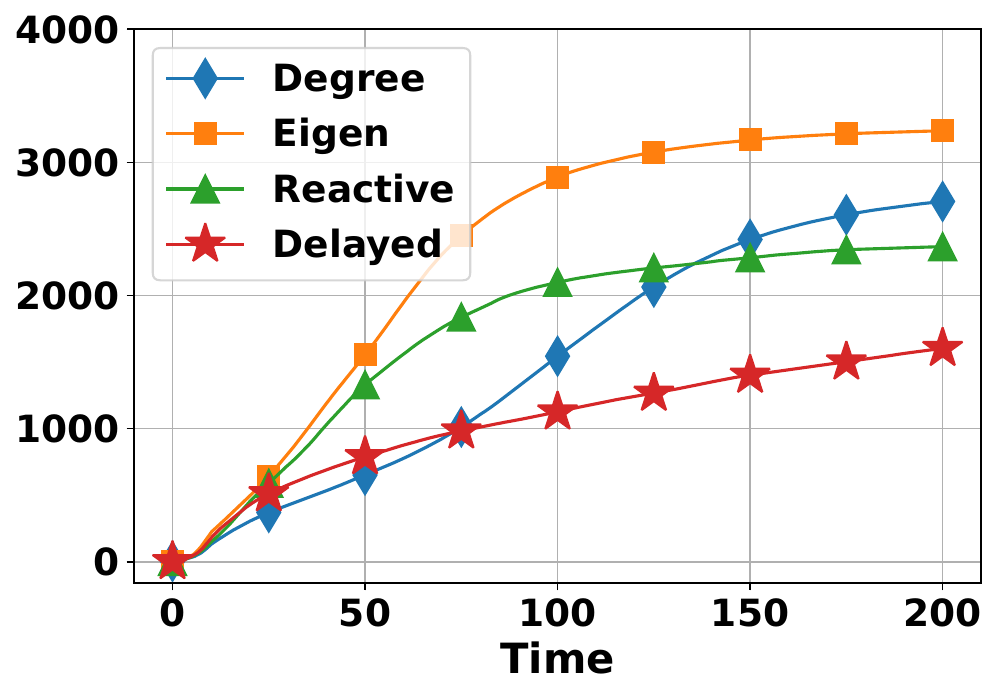}%
  }%
  \subfigure[$T=15,\ b=20\%$]{%
    \includegraphics[width=0.25\linewidth]{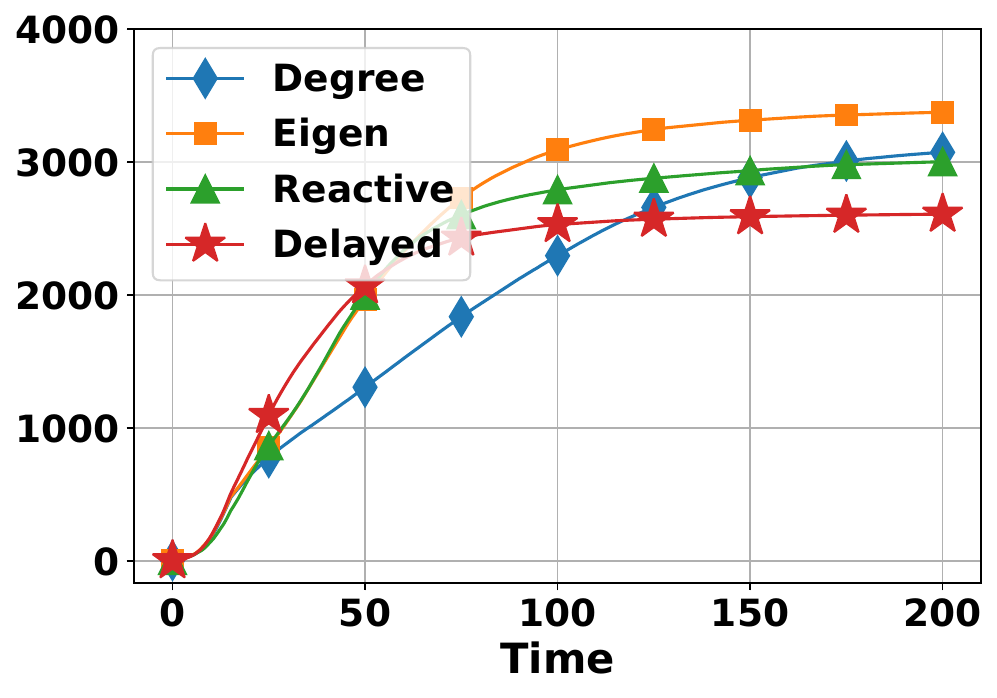}%
  }%
  \subfigure[$T=20,\ b=20\%$]{%
    \includegraphics[width=0.25\linewidth]{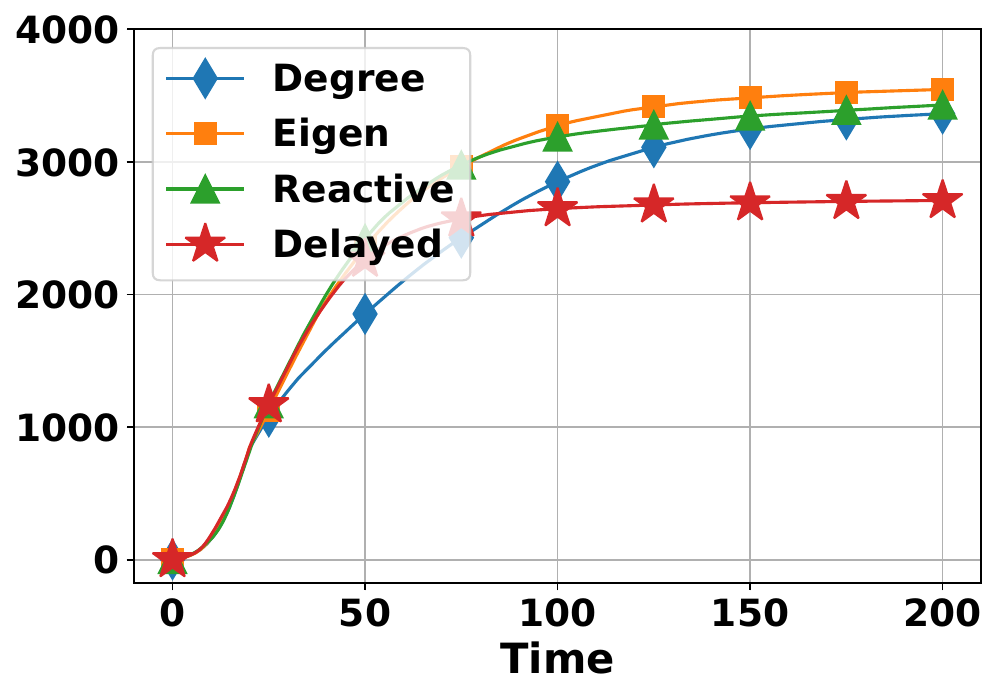}%
  }%

  \vspace{-2mm}
  \caption{The expected number of infected nodes on the Facebook graph with varying values of $T$ and $b$.}
  \label{fig:facebook}
  \vspace{-4mm}
\end{figure*}

In this section, we present simulation results to demonstrate the efficacy of our proposed patching policy (Delayed) in mitigating malware propagation against three existing baselines. The first baseline is an eigenvector centrality‐based policy (Eigen), which immunizes the top-$K$ nodes by the eigenvector centrality (obtained from the original, unweighted adjacency matrix). This policy has been considered in the literature~\cite{Canright06,carreras2007eigenvector} as the eigenvector centrality can capture the early growth rate of the SI dynamics and thus be used to identify the most vulnerable nodes for vaccination. The second baseline is a degree‐based policy (Degree), vaccinating the top-$K$ high-degree nodes, since high-degree nodes can be more likely exposed to the spread of an epidemic. The last one is a reactive policy (Reactive) proposed in~\cite{lee2019transient}, which computes each node’s predicted infection probability $\hat{x}_i(T)$ using \eqref{upper-bound} and vaccinates the top-$K$ high‐risk nodes. 

\subsection{Simulation Setup}

For performance evaluation, we consider both synthetic and real-world graphs. For the former, we use the stochastic block model~\cite{holland1983stochastic} to generate synthetic random graphs, each of which contains $k$ communities and has more edges within communities than between communities. The average degree is fixed at eight for all random graphs. We vary the number of nodes $n \!=\! 1000, 2000, 4000, 8000$ and the number of communities $k \!=\! 3, 4, 5, 6$. For the real-world graph, we use the Facebook social network~\cite{jure2014snap}, which has 4,039 nodes and 88,234 edges collected via a Facebook app and represents social interactions among the nodes. 

We set up each simulation as follows. At $t \!=\! 0$, we initiate an epidemic by randomly choosing a small number of nodes as initially infected nodes (or infection sources). We use a single source for small graphs ($n \!=\! 1000, 2000$) and use five sources for the larger graphs. In each simulation, each patched node still remains vulnerable to infection for the patching delay $T$, so it becomes successfully patched or vaccinated at time $T$ only if it has not been infected by $T$. In other words, at time $t \!=\! T$, if a patched node remains healthy, it is considered successfully patched and removed from the graph. However, if it has been infected during time $T$, although it was patched initially, it remains infected in the graph and spreads the malware. We set the infection rate $\beta\!=\!0.01$ for all experiments. We vary the values of the patching delay $T$ and the budget constraint on the number of available vaccines or patches, denoted as $b$, to see their impact on the performance. Each data point reported here is obtained by averaging over $100$ independent simulations. We only report representative simulation results due to space constraints.

\subsection{Simulation Results}

\noindent \textbf{Synthetic graphs.} Figure \ref{fig:sbm} shows the expected number of infected nodes under each vaccination policy, obtained on synthetic graphs with $n\!=\!1000,2000,3000,4000$ for $k\!=\!3,4,5,6$, respectively, while varying the value of $T\!=\!15,20,25,30$. In all cases, we set the budget constraint $b$ to 20\% of the graph size (the number of nodes in the graph). As shown in Figure \ref{fig:sbm}, our Delayed policy exhibits the overall most effective performance in minimizing the number of infected nodes. The Reactive policy shows competitive results compared to our Delayed policy when the patching delay $T$ is small for large graphs, in which case the nodes chosen by the policy for patching are most likely patched or vaccinated successfully after $T$. However, when the value of $T$ grows, our Delayed policy becomes much more effective. The expected numbers of infected nodes under Delayed policy can be only up to half those under Reactive policy (see Figures~\ref{fig:sbmk5t25} and~\ref{fig:sbmk5t30}). On the other hand, both the Eigen and Degree policies lead to unsatisfactory results, especially for larger graphs with a higher value of $k$ (having more communities). This is because their choice of nodes for patching is determined merely based on the underlying graph structure, but not the dynamics of malware propagation as well as the presence of the patching delay $T$. Specifically, they leave up to six times more infections than the other policies, which happens when $n\!=\!4000,8000$ and $k\!=\!5,6$, respectively.

In Figure \ref{fig:num_infected_sbm}, we further demonstrate the superior performance of our Delayed policy compared to the baselines for a large synthetic graph with $n\!=\!8000$. Here, we measure the expected number of infected nodes at $t\!=\!1000$. To show the robustness of our policy, we vary the number of communities $k\!=\!3,4,5,6$, the budget constraint $b\!=\!10\%,15\%,20\%,30\%$ and the patching delay $T\!=\!25,30,35,40$. As shown in Figure \ref{fig:num_infected_sbm}, Delayed policy consistently outperforms the other baselines by a large margin, having the least number of infected nodes at $t \!=\! 1000$. In particular, with the low budget ($b\!=\!10\%$), Delayed policy already halves the number of infected nodes compared to Reactive policy, and it reduces the number of infections by nearly five times compared to Degree and Eigen policies as budget $b$ increases. Furthermore, even when the patching delay $T$ increases, the performance of Delayed policy remains relatively consistent, especially compared to Reactive policy. This demonstrates the effectiveness of its delayed patching.

\vspace{1mm}
\noindent \textbf{Real-world graph.} We next evaluate the performance of our Delayed policy on the Facebook social network to confirm its superiority over the three baselines. As shown in Figure \ref{fig:facebook}, our Delayed policy shows the best performance. It always produces the lowest infection curve, meaning that we can always obtain the least expected number of infected nodes with this policy. With the patching delay $T\!=\!15$ or $T \!=\! 20$, Delayed policy reduces the number of infected nodes by at least 500 nodes compared to Reactive and the other baselines. Furthermore, with $T \!=\! 20$, while the baseline policies are mostly ineffective as almost the entire network becomes infected, the Delayed policy is still able to save $25\%$ to $35\%$ of the total population with budget $b \!=\! 10\%$ and $b \!=\! 20\%$, respectively. Interestingly, the Degree‐based policy initially outperforms the Delayed policy in the early stages, but it quickly becomes ineffective, leaving a much higher infection count in the end.

To summarize, the simulations on both synthetic and real-world graphs confirm the effectiveness of our patching policy, especially for longer patching delays $T$. This is crucial since the patching process or vaccines take time to become effective in practice. By accurately identifying the boundary between infected and healthy regions in the presence of the patching delay $T$, our framework is able to isolate high-risk nodes and prevent the malware from spreading through the network.

\section{Conclusion}
In this work, we introduced a novel mathematical framework for effective patching under limited patching resources and in the presence of patching delay. The rationale behind this framework is to identify a minimum-cut boundary that separates the most likely infected nodes from the healthy region  and leverage the boundary to identify which nodes to patch under limited patching resources. We demonstrated the superior performance of its resulting patching policy over the existing baseline policies through extensive experiments on synthetic and real-world networks. We believe that our work provides a first step toward the design of vaccination strategies for general networks under realistic delay and resource constraints.\label{sec:conclusion}

\vspace{-1mm}
\section*{Acknowledgment}

We thank Srinivas Tenneti for his earlier contributions to this work. This work was supported by the National Science Foundation under Grant Nos. 1910749, 2007423, 2209921, and 2209922, and the International Energy Joint R\&D Program of the Korea Institute of Energy Technology Evaluation and Planning (KETEP), granted financial resource from the Ministry of Trade, Industry \& Energy, Republic of Korea (No. 20228530050030). C. Lee is the corresponding author.

\bibliographystyle{IEEEtran}
\bibliography{refs}

\end{document}